\documentclass[12pt]{article}
\usepackage{lscape}
\usepackage{amsfonts,ulem,url}
\usepackage[margin=.9in]{geometry}
\usepackage{XCharter} 
\usepackage{xr-hyper}
\usepackage{amssymb}
\usepackage{amsmath}
\usepackage{amstext,verbatim,graphicx,ifthen,multirow,amsthm,algorithm2e, mathtools, array}
\usepackage{bm}
\usepackage[round]{natbib}
\usepackage[bf]{caption}
\usepackage{physics}
\usepackage{setspace}
\usepackage{enumitem}
\usepackage{placeins}
\usepackage{xcolor}
\usepackage{rotating}
\usepackage{adjustbox}
\usepackage{appendix}

\usepackage{float}

\newtheorem{assumption}{Assumption}[section]
\newtheorem{theorem}{Theorem}[section]

\newtheorem{prop}{Proposition}[section]
\newtheorem{corollary}{Corollary}[section]

\theoremstyle{definition}

\newtheorem{remark}{Remark}[section]

\numberwithin{equation}{section}

\usepackage{apptools}
\AtAppendix{\counterwithin{lemma}{section}}
\AtAppendix{\counterwithin{theorem}{section}}
\AtAppendix{\counterwithin{corollary}{section}}

\DeclareMathOperator*{\argmin}{arg\,min}

\graphicspath{{./figures/}}

\usepackage[colorlinks=true,unicode,bookmarksnumbered=false, hyperfootnotes=true]{hyperref}
\hypersetup{citecolor = blue}

\usepackage[colorinlistoftodos]{todonotes}

\newcommand{\indep}{\perp\!\!\!\perp}

\usepackage{physics}
\usepackage{rotating}
\usepackage{adjustbox,algorithm2e}

\title{Sequential Synthetic Difference in Differences\thanks{{\small We thank Kirill Borusayk for his feedback, which has greatly improved the paper. We also thank seminar participants at CEMFI, Harvard, IMT Lucca, Erasmus, Groningen, Northwestern, NYU, BU, and Duke, as well as conference participants at the KU Leuven Summer Event and ASSA 2025. Aleksei Samkov acknowledges ﬁnancial support from the Maria de Maeztu Unit of Excellence CEMFI MDM-2016-0684, funded by MCIN/AEI/10.13039/501100011033, and CEMFI.}}} 
\author{Dmitry  Arkhangelsky \thanks{{\small  Associate Professor, CEMFI, CEPR, darkhangel@cemfi.es. }} \and Aleksei Samkov \thanks{{\small PhD student, CEMFI, aleksei.samkov@cemfi.edu.es }} }
\date{\today}

\begin{document}

\maketitle
\begin{abstract}
\singlespacing
We propose the Sequential Synthetic Difference-in-Differences (Sequential SDiD) estimator for event studies with staggered treatment adoption, particularly when the parallel trends assumption fails. The method uses an iterative imputation procedure on aggregated data, where estimates for early-adopting cohorts are used to construct counterfactuals for later ones. We prove the estimator is asymptotically equivalent to an infeasible oracle OLS estimator within a linear model with interactive fixed effects. This key theoretical result provides a foundation for standard inference by establishing asymptotic normality and clarifying the estimator's efficiency. By offering a robust and transparent method with formal statistical guarantees, Sequential SDiD is a powerful alternative to conventional difference-in-differences strategies.
\end{abstract}

\noindent \textbf{Keywords}: event studies, synthetic control, difference in differences, interactive fixed effects, panel data, sequential analysis.

\noindent \textbf{JEL Codes}: C21, C23, C38.

\newpage
	
\section{Introduction}

Event study designs, where a researcher observes units before and after an event alongside a comparison group, are a cornerstone of modern applied economics \citep{currie2020technology,goldsmith2024tracking}. These studies routinely employ the difference-in-differences (DiD) strategy \citep{angrist2008mostly,Bertrand2004did,card1990impact,card1994minimum}. The credibility of DiD, however, hinges on a parallel trends assumption, requiring that counterfactual outcomes for the treated group would have evolved in parallel with the outcomes of the comparison group. In the common case of staggered treatment adoption, most modern estimators, while flexible in accommodating heterogeneous treatment effects, still rely on this fundamental assumption \citep{borusyak2024revisiting,CallawaySantAnna,de2020two,sun2020estimating}.

We address this limitation by proposing the Sequential Synthetic Difference-in-Differences (Sequential SDiD) estimator, a method for staggered adoption designs that is robust to violations of the parallel trends assumption. Our proposal adapts the Synthetic DiD estimator of \cite{arkhangelsky2021synthetic}, which combines ideas from the Synthetic Control (SC) and DiD literature \citep{abadie2003,abadie2010synthetic}. The procedure is iterative: it first averages outcomes for units sharing the same adoption date, then sequentially estimates treatment effects. After each estimation step, it uses the result to impute the missing counterfactual for the treated cohort. This imputed data is then used in the analysis of subsequent cohorts, preventing bias from cascading through the estimation.

We analyze Sequential SDiD within a linear model with interactive fixed effects (IFE), which explicitly models the unobserved confounders that drive violations of parallel trends \citep{arellano2011identifying,bai2009panel,chamberlain1992efficiency,Freyberger,holtz1988estimating,pesaran2006estimation}, and has been routinely used to establish statistical properties for the SC-type estimators \citep{abadie2010synthetic,arkhangelsky2021synthetic,ben2021augmented,ben2022synthetic,ferman2019synthetic,ferman2021synthetic}. Our analysis departs from much of the SC literature by employing an asymptotic regime with a fixed number of periods and a large number of units. This allows us to connect the SC literature to the classic moment-based econometrics panel data literature (going back at least to \citealp{chamberlain1984panel}). This framework is suitable for our setting because the initial averaging of outcomes within large cohorts effectively reduces the noise level in the problem, connecting our work to the "low-noise" regime in the SC literature \citep{Hirshberg}.

Our analysis is built around an infeasible "oracle" OLS estimator. This oracle knows the unobserved interactive fixed effects and uses them as regressors---a benchmark that captures what applied researchers attempt to do with proxies like unit-specific trends. We prove that our feasible Sequential SDiD estimator is asymptotically equivalent to this infeasible oracle. This equivalence is the paper's key theoretical result, as it immediately implies that our estimator is asymptotically normal and unbiased, providing a basis for standard inference. It also delivers, to our knowledge, the first formal efficiency guarantees for an SC-type method, showing that Sequential SDiD is asymptotically as efficient as the oracle OLS estimator. 

A key intermediate step in our proof, which is of independent interest, is that we derive an alternative representation for this oracle regression. We show that it can be implemented via a sequential algorithm, where at each step the oracle constructs a weighted DiD estimator and uses the result for imputation. This finding provides new insight into the mechanics of modern imputation estimators, such as that of \cite{borusyak2024revisiting} for the two-way fixed effects model. More broadly, this sequential representation of a correctly specified oracle opens up possibilities for developing other new methods that relax standard DiD assumptions. Our work focuses on one such relaxation (allowing for IFE), but other routes, such as relaxing selection assumptions, could be built on a similar sequential logic.

Our work offers contributions to several strands of the applied and theoretical econometrics literature.

First, we introduce a new, transparent, and robust estimator for event studies with staggered adoption. A key feature is its sequential algorithm, where estimates for early-adopting cohorts are used to impute counterfactuals that, in turn, inform the analysis of later ones. This iterative structure provides a cohesive analysis of the entire event study panel, distinguishing it from related SC approaches for staggered designs that effectively treat different adoption cohorts as separate problems (e.g., \citealp{ben2022synthetic, cattaneo2021prediction}).

Second, we provide an answer to a long-standing critique of SC methods. A common question is why one should rely on synthetic control weights instead of directly estimating the underlying factor model that motivates the procedure. Our results resolve this tension by formally demonstrating that our SC-based method is asymptotically equivalent to an oracle OLS estimator that directly uses the unobserved factors. This clarifies that, in this setting, the choice is not between balancing and direct estimation, but rather how to feasibly approximate the same ideal oracle benchmark. Crucially, this equivalence allows us to establish efficiency guarantees for an SC-type method. These results complement the optimal regret properties of the SC method under adversarial sampling derived by \cite{chen2023synthetic}.

Finally, we advance the literature on difference-in-differences with staggered adoption. While recent methods have addressed the issues with heterogeneous treatment effects, they still effectively operate within a standard two-way fixed effects model that implies the parallel trends assumption. Our estimator also accommodates unrestricted treatment effect heterogeneity, but it remains robust when the parallel trends assumption is violated due to the presence of unobserved interactive fixed effects.

We demonstrate our method's performance with an empirical application and two data-based simulation studies. First, we revisit the analysis of Community Health Centers by \cite{bailey2015war} and show that Sequential SDiD produces results that are reassuringly similar to standard DiD in a setting where the parallel trends assumption is believed to hold. We then use two sets of simulations to assess performance when this assumption is violated by confounding interactive fixed effects. The first is calibrated to the county-level data from the application. The second, which extends the simulation design of \cite{arkhangelsky2021synthetic} to the staggered case and is inspired by the seminal work of \cite{Bertrand2004did}, uses a state-level wage panel where treatment assignment is correlated with the underlying factors. In both simulation designs, we find that while standard DiD becomes severely biased, our Sequential SDiD estimator provides reliable estimates and valid inference.

Our analysis has two main limitations. First, our theoretical framework requires that adoption cohorts are relatively large. This size is crucial as it allows us to leverage the law of large numbers; by averaging outcomes within a cohort, we can substantially reduce the influence of idiosyncratic noise. This approach makes our statistical results possible under otherwise mild assumptions and provides a reasonable approximation for many empirical settings. It is also commonly used in the modern DiD literature (e.g., \citealp{abadie,callaway2020difference}).

Second, our main results rely on the assumption that idiosyncratic errors are independent across units, a condition that ensures they concentrate around zero upon aggregation. The appropriateness of this assumption depends on the context. While some influential work has relied on it for inference \citep{card1994minimum}, other research highlights that common shocks can induce error correlation that survives aggregation \citep{Bertrand2004did}. Our model is designed to address this concern. The interactive fixed effects capture precisely these common shocks, and recent evidence suggests that this structure can account for the vast majority of variation in aggregated data, leaving a much smaller, plausibly independent residual error \citep{arkhangelsky2021synthetic}. Nonetheless, in applications where this separation is not clean---for instance, where idiosyncratic shocks themselves have heavy tails and drive aggregate outcomes---our statistical guarantees may not apply.\footnote{This type of granularity, where idiosyncratic shocks affect aggregates, is documented by \cite{gabaix2011granular} in the context of the US economy.}

The rest of the paper proceeds as follows. Section \ref{sec:method} presents the econometric setup and our estimator. Section \ref{sec:theory} establishes the theoretical results. Section \ref{sec:cov} discusses covariates. Section \ref{sec:exp} contains our empirical illustration and simulation experiments, and Section \ref{sec:conc} concludes.

\paragraph{Notation:} We use standard notation for expectations and variance operators, $\mathbb{E}[\cdot]$ and $\mathbb{V}[\cdot]$, respectively. For a sequence of random variables $X_n, Y_n$ we write $X_n = o_p(Y_n)$ if $\frac{X_n}{Y_n}$ converges to zero in probability.  For two sequences $a_n$ and $b_n$ we write $a_n \lesssim b_n$ if $\frac{a_n}{b_n}$ is bounded, and $a_n \ll b_n$ if $\frac{a_n}{b_n}$ converges to zero. We write $a_n\sim b_n$ if $a_n \lesssim b_n$ and $b_n \lesssim a_n$. We use the same notation with subscript $p$ for the corresponding concepts for random sequences.

\section{Methodology}\label{sec:method}

This section lays out our proposed methodology. We begin by describing the econometric framework, including our assumptions about potential outcomes and the data-generating process with interactive fixed effects. We then formally present the Sequential Synthetic Difference-in-Differences (Sequential SDiD) estimator. We conclude by outlining procedures for statistical inference and model validation.
\subsection{Setup}\label{subsec:setup}

We observe $n$ units over $T$ periods, with $i$ being a generic unit and $t$ being a generic period. In our theoretical analysis, we treat $T$ as fixed and $n$ as going to infinity -- the asymptotic regime that provides a reasonable approximation for a large class of empirical applications. For each unit $i$ and period $t$, we observe a real-valued outcome $Y_{i,t}\in \mathbb{R}$ and a binary treatment indicator $W_{i,t}\in \{0,1\}$. Following most of the applied work, we focus on settings with staggered adoption, thus assuming $W_{i,t} \ge W_{i,t-1}$.

We formalize causality by interpreting the observed outcomes using potential outcomes \citep{neyman1923,rubin1974estimating,imbens2015causal}.
\begin{assumption}[No-anticipation]\label{as:potential outcomes}
For each $i$ and $t$, there exists a (potentially random) function $Y_{i,t}(\cdot):\{0,1\}^{t} \rightarrow \mathbb{R}$ such that
\begin{equation*}
Y_{i,t} = Y_{i,t}(\boldsymbol{W}_{i}^t),
\end{equation*}
where $\boldsymbol{W}_{i}^t:= (W_{i,1},\dots, W_{i,t})$.
\end{assumption}
This assumption incorporates two separate restrictions. The first is no anticipation: only treatments realized by period $t$ can affect the outcomes in that period. The second is the absence of cross-unit spillovers, a key component of the Stable Unit Treatment Value Assumption (SUTVA), meaning potential outcomes for unit $i$ only vary with that unit's own treatment assignment. See \cite{arkhangelsky2024causal} for a discussion of these assumptions.

Given our focus on staggered adoption designs, we re-index potential outcomes by adoption time. We begin by defining the set of all possible staggered treatment paths, $\mathbb{W}:= \{\mathbf{w}\in \{0,1\}^T: w_t\ge w_{t-1}\}$. For any such path $\mathbf{w} \in \mathbb{W}$, we can define its corresponding adoption time as $a(\mathbf{w}):= \inf \{t: w_t =1\}$. This mapping creates a one-to-one correspondence between treatment paths in $\mathbb{W}$ and adoption times in the set $\mathbb{A}:= \{1,\dots, T, +\infty\}$. This allows us to define potential outcomes indexed by adoption time, $Y_{i,t}(a) := Y_{i,t}(\mathbf{w}^t(a))$, where $\mathbf{w}^t(a)$ is the history of the unique treatment path corresponding to adoption time $a$. For each unit $i$, we denote its observed adoption time by $A_i := \inf\{t \le T: W_{i,t} = 1\}$. The observed outcome can then be written as a function of the potential outcome corresponding to the observed adoption time: $Y_{i,t} = Y_{i,t}(A_i)$. The internal consistency of this representation relies on Assumption \ref{as:potential outcomes}. In what follows, we use both representations interchangeably.

Our next assumption specifies the data-generating process for the potential outcomes.
We model potential outcomes as a combination of four components: (i) a standard unit fixed effect, $\alpha_i$; (ii) a standard time fixed effect, $\beta_t$; (iii) a low-rank interactive fixed effect, $\theta_i^\top \psi_t$, which captures unobserved confounding factors that violate the parallel trends assumption; and (iv) the treatment effect itself, $\tau_{i,a,k}$, which is allowed to be heterogeneous across units, adoption cohorts, and time since treatment.
\begin{assumption}[Interactive Fixed Effects]\label{as:factor}
For all $i$ and $t$, the potential outcomes are given by
\begin{equation}
Y_{i,t}(A_i) = \alpha_i + \beta_t + \theta_i^\top \psi_t + \sum_{k\ge 0}\tau_{i,a,k}\mathbf{1}\{A_i = a,k = t-A_i\} + \epsilon_{i,t},
\end{equation}
where $\theta_i\in \mathbb{R}^{r}$ for some $r\ge 0$, $\mathbb{E}[\epsilon_{i,t}|\{A_i\}_{i=1}^n, \boldsymbol{\gamma}] = 0$, and the error vectors $\boldsymbol{\epsilon}_i:= (\epsilon_{i1},\dots, \epsilon_{i,T})$ are independent over $i$ conditionally on $\left(\{A_i\}_{i=1}^n, \boldsymbol{\gamma}\right)$, where $\boldsymbol{\gamma}:=\{\alpha_i, \theta_i,\beta_t, \psi_t,\tau_{i,a,k}\}_{i,t,a,k}$.
\end{assumption}
The interpretation of Assumption \ref{as:factor} depends on the underlying sampling scheme and treatment assignment protocol. We now discuss two scenarios that justify this assumption and clarify the meaning of the treatment effect parameters $\tau_{i,a,k}$.

\paragraph{Example 1: Latent Unconfoundedness.} Suppose that $\{Y_i, A_i\}_{i=1}^n$ are $n$ i.i.d. draws from a population. For each unit, suppose there is a latent characteristic $U_i$ such that a conditional independence (unconfoundedness) assumption holds:
\begin{equation}\label{eq:lat_unc}
 \{Y_{i}(\mathbf{w})\}_{\mathbf{w} \in\mathbb{W}} \indep A_i\, |\, U_i.
\end{equation}
Further, suppose the never-treated potential outcomes satisfy the IFE structure:
\begin{equation*}
Y_{i,t}(+\infty)= \alpha(U_i) + \beta_t + \theta^{\top}(U_i)\psi_t +\epsilon_{i,t}(+\infty), \quad \text{with} \quad \mathbb{E}[\epsilon_{i,t}(+\infty)|U_i] = 0.
\end{equation*}
We can then define the treatment effect parameter as the average causal effect conditional on $U_i$, and define the error term $\epsilon_{i,t}$ as the residual component:
\begin{align*}
    \tau_{i,a,k} &:= \mathbb{E}[Y_{i,a+k}(a) - Y_{i,a+k}(+\infty)| U_i],\\
    \epsilon_{i,t} &:= \epsilon_{i,t}(+\infty) + \left(Y_{i,t}(A_i) - Y_{i,t}(+\infty) - \tau_{i,A_i,t-A_i}\right).
\end{align*}
Under these definitions, the model for the observed outcome $Y_{i,t}$ conforms to the structure in Assumption \ref{as:factor}.

Independent sampling is standard in the panel data literature and is natural for environments where units of observation correspond to economic agents, such as individuals or firms. It is also common in the theoretical DiD literature (e.g., \citealp{abadie2005semiparametric, callaway2020difference, sant2020doubly}). The latent unconfoundedness restriction \eqref{eq:lat_unc} is a form of strict exogeneity: conditional on $U_i$, the adoption date is as good as random. Strict exogeneity has a long tradition in panel data econometrics \citep{chamberlain1984panel} and underlies analyses based on parallel trends assumptions (see \citealp{ghanem2022selection} for a discussion).

\paragraph{Example 2: Aggregate and Idiosyncratic Shocks.} Consider a fixed set of $n$ units over $T$ periods. In each period $t$, unit $i$ is exposed to an unobserved aggregate shock $H_t$ and an idiosyncratic shock $\nu_{i,t}$. Suppose the potential outcomes have the following structure:
\begin{equation}\label{eq:fin}
    Y_{i,t}(a,\boldsymbol{h}^t, \boldsymbol{u}^t) = \sum_{k \ge 0}\tau_{i,a,k}(\boldsymbol{h}^t,\boldsymbol{u}^t)\mathbf{1}\{t -k = a\} + f_{i,t}(\boldsymbol{u}^t, \boldsymbol{h}^t),
\end{equation}
where $\boldsymbol{h}^{t}$ and $\boldsymbol{u}^t$ are the histories of aggregate and idiosyncratic shocks. This decomposition is general; the term $\tau_{i,a,k}(\cdot)$ is simply the difference $Y_{i,a+k}(a, \cdot) - Y_{i,a+k}(+\infty, \cdot)$, implying that shocks do not restrict the causal effects.

Let $\boldsymbol{\nu}_i$ and $\boldsymbol{H}$ be the realized shock vectors. We make two independence assumptions: (i) idiosyncratic shocks are independent of adoption times conditional on aggregate shocks, and (ii) unit-level data is independent across units conditional on aggregate shocks:
\begin{gather*}
    \{\boldsymbol{\nu}_i\}_{i=1}^n \indep \{A_i\}_{i=1}^n | \boldsymbol{H}; \\
    (\boldsymbol{\nu}_1, A_1) \indep \dots \indep (\boldsymbol{\nu}_n, A_n) | \boldsymbol{H}.
\end{gather*}
Furthermore, suppose the conditional expectation of the untreated component has an IFE structure:
\begin{equation*}
    \mathbb{E}\left[f_{i,t}(\boldsymbol{\nu}^t_{i}, \boldsymbol{H}^t)|\boldsymbol{H}\right] = \alpha_i + \beta_t(\boldsymbol{H}) + \theta_i^\top\psi_t(\boldsymbol{H}).
\end{equation*}
Finally, we define the parameters and the error term by conditioning on the aggregate shocks $\boldsymbol{H}$:
\begin{align*}
Y_{i,t}(a) &:= Y_{i,t}(a,\boldsymbol{\nu}^t_{i}, \boldsymbol{H}^t), \\
\tau_{i,a,k} &:= \mathbb{E}[\tau_{i,a,k}(\boldsymbol{\nu}^t_{i}, \boldsymbol{H}^t)|\boldsymbol{H} ], \\
\epsilon_{i,t} &:= f_{i,t}(\boldsymbol{\nu}_{i}^t, \boldsymbol{H}^t) - \mathbb{E}\left[f_{i,t}(\boldsymbol{\nu}_{i}^t, \boldsymbol{H}^t)|\boldsymbol{H}\right] \\
&\quad + \left(\tau_{i,a,k}(\boldsymbol{\nu}_{i}^t, \boldsymbol{H}^t) - \tau_{i,a,k}\right)\mathbf{1}\{A_i = a, t -k = a\}.
\end{align*}
The resulting model for $Y_{i,t}$ again satisfies Assumptions \ref{as:potential outcomes} and \ref{as:factor}.

This second example may seem less familiar, but it captures features common in applications where aggregate shocks drive outcomes and treatment assignment. First, it allows treatment trajectories to be correlated across units. Second, it permits the errors $\epsilon_{i,t}$ to have a factor structure (i.e., be correlated), since they are only independent ``conditional'' on $\boldsymbol{H}$. This structure is prevalent in large panel data models (e.g., \citealp{bai2009panel,pesaran2006estimation}) and is related to research on shift-share designs \citep{adao2019shift}.

In both examples, the parameter of interest $\tau_{i,a,k}$ represents a specific conditional average treatment effect. In Example 1, we condition on the latent characteristic $U_i$ that drives selection. In Example 2, we average over idiosyncratic shocks $\boldsymbol{\nu}_i$ while conditioning on aggregate shocks $\boldsymbol{H}$. Our analysis will focus on estimating averages of $\tau_{i,a,k}$ for subpopulations of adopters. These examples illustrate that the interpretation of these averages depends on the underlying probability model.

\subsection{Estimator}
In this section, we introduce the new estimator, which we call Sequential SDiD. As the name suggests, it is based on sequential application of a version of the SDiD estimator introduced in \citep{arkhangelsky2021synthetic}. The key difference, though, is that we apply SDiD principles to aggregated data. Let $\mathcal{A}$ be the support of $A_i$; for each adoption cohort $a \in \mathcal{A}$, we define aggregate outcomes:
\begin{equation*}
    Y_{a,t} := \frac{\sum_{i : A_i = a} Y_{i,t}}{n_a},
\end{equation*}
where $n_a := \sum_{i=1}^n \mathbf{1}\{A_i = a\}$ is the total number of units in cohort $a$. We also define the corresponding shares, $\pi_a := \frac{n_a}{n}$. Assumption \ref{as:factor} guarantees that these aggregate outcomes follow a similar interactive fixed effects model:
\begin{equation}\label{eq:agg_eq}
    Y_{a,t} = \alpha_a + \beta_t + \theta_a^\top \psi_t +  \mathbf{1}\{a \le t\}\tau_{a,t-a} + \epsilon_{a,t},
\end{equation}
where $\epsilon_{a,t} :=\frac{\sum_{i : A_i = a} \epsilon_{i,t}}{n_a}$, and other variables are the cohort-level averages of their unit-level counterparts. Representation \eqref{eq:agg_eq} is the key to our algorithm, which we present formally in Algorithm \ref{alg:ssdid}.

\RestyleAlgo{boxruled}
\LinesNumbered
\begin{algorithm}[t]
 \KwData{Aggregated data $\{Y_{a,t}, \pi_a\}$, parameters $a_{\min}$, $a_{\max}$, $K$, $\eta$}
 \KwResult{Estimates $\{\hat \tau_{a,k}^{SSDiD}\}_{a \in \{a_{\min},\dots, a_{\max}\}, k \in \{0,\dots, K\}}$}
 \For{$k \in \{0, \dots, K\}$}{
  \For{ $a \in \{a_{\min},\dots, a_{\max}\}$}{
   Construct weights by solving:
    \begin{equation*}
    \begin{aligned}
        \hat\omega^{(a,k)} :=& \arg\min_{\sum_{j >a} \omega_j = 1} \left\{\sum_{l < a+k}\left(\sum_{j >a} \omega_j Y_{j,l} + \omega_0 - Y_{a,l}\right)^2 + \eta^2\sum_{j > a} \frac{\omega_j^2}{ \pi_j}\right\},\\
        \hat\lambda^{(a,k)} :=& \arg\min_{\sum_{l <a+k} \lambda_l = 1} \left\{\sum_{j >a}\left(\sum_{l <a+k} \lambda_l Y_{j,l} + \lambda_0 - Y_{j,a+k}\right)^2 + \eta^2\sum_{l <a+k} \lambda_l^2 \right\};
    \end{aligned}
    \end{equation*}
    \nl Construct the estimator using a weighted double-difference:
    \begin{equation*}
          \hat \tau_{a,k}^{SSDiD} := \left(Y_{a,a+k} - \sum_{j>a}\hat \omega_j^{(a,k)}Y_{j,a+k}\right) -  \sum_{l <a+k}\hat \lambda_l^{(a,k)}\left(Y_{a,l} - \sum_{j>a}\hat \omega_j^{(a,k)}Y_{j,l}\right);
    \end{equation*}
    \nl Impute the counterfactual outcome for the next iteration:
    \begin{equation*}
        Y_{a,a+k} := Y_{a,a +k} - \hat \tau_{a,k}^{SSDiD};
    \end{equation*}
  }
 }
\caption{Sequential SDiD}\label{alg:ssdid}
\end{algorithm}

Algorithm \ref{alg:ssdid} estimates the treatment effects $\tau_{a,k}$ sequentially, starting with the contemporaneous effect ($k=0$) and proceeding to longer horizons. For each cohort $a$ and horizon $k$, the procedure applies the principles of the SDiD estimator. It first constructs unit-specific weights $\hat{\omega}^{(a,k)}$ and time-specific weights $\hat{\lambda}^{(a,k)}$ (Line 3). The unit weights create a synthetic control for cohort $a$ using a weighted average of not-yet-treated cohorts ($j>a$). The time weights create a synthetic match for the event period $a+k$ using a weighted average of pre-event periods ($l < a+k$). A key difference from the original SDiD estimator is that we do not impose non-negativity constraints on these weights. The estimator $\hat{\tau}_{a,k}^{SSDiD}$ is then formed using these weights in a generalized DiD formula (Line 4).

The sequential nature of our method is driven by the imputation step in Line 5. After estimating $\hat{\tau}_{a,k}^{SSDiD}$, we use it to update the outcome $Y_{a,a+k}$. This step replaces the observed outcome with an estimate of the missing average counterfactual outcome.
The imputation in Line 5 is the engine of the sequential procedure. When we move to estimate a longer-run effect for cohort $a$ (i.e., for $k' > k$), or any effect for a later cohort $a' > a$, the period $a+k$ becomes part of the pre-treatment history. However, the observed outcome $Y_{a,a+k}$ is contaminated by treatment. The imputation step $Y_{a,a+k} := Y_{a,a+k} - \hat \tau_{a,k}^{SSDiD}$ replaces this treated outcome with its estimated untreated counterfactual. This ensures that the weights computed in subsequent steps are based on a panel where the parallel trends assumption is enforced on the best available estimate of the counterfactual data, preventing the cascading bias that would arise from using raw treated outcomes as controls.
This imputed value is then carried forward and used to construct weights and estimates for longer horizons (i.e., for $k' > k$). This process ensures that information from early estimates is efficiently incorporated into later ones.

The algorithm depends on three user-specified parameters: the regularization parameter $\eta$, the maximum horizon $K$, and the range of cohorts $(a_{\min}, a_{\max})$. The parameter $\eta \ge 0$ controls the penalty on the weights; its role is discussed further in our theoretical analysis.
In practice, it serves to prevent the synthetic control from overfitting to idiosyncratic noise in the pre-treatment periods, especially when the number of control units or pre-treatment periods is small.
The choice of $K$ and $(a_{\min}, a_{\max})$ determines which treatment effects are estimated and must satisfy the data constraint $a_{\max} +K \le T$, since estimating $\hat \tau_{a_{\max}, K}$ requires observing outcomes in period $a_{\max} + K$.

We use the estimated cohort-specific effects to construct an average effect for each horizon $k$:
\begin{equation}\label{eq:agg_est}
 \hat \tau_k^{SSDiD}(\mu) = \sum_{a \in \{a_{\min},\dots, a_{\max}\}} \mu_{a} \hat \tau_{a,k}^{SSDiD},
\end{equation}
where the weights $\mu$ are user-specified. In our analysis, we focus on weights that are proportional to the cohort shares, $\mu_{a} = \frac{\pi_a}{\sum_{a' \in  \{a_{\min},\dots, a_{\max}\}}\pi_{a'}}$, and use $\hat \tau_k^{SSDiD}$ to denote the resulting estimator.

\begin{remark}
Algorithm \ref{alg:ssdid} constructs $K+1$ estimates for each adoption time in the range $[a_{\min}, a_{\max}]$. While one could, in principle, allow the horizon $K$ to vary by cohort (e.g., $K_a$), keeping it fixed has two advantages. First, it ensures that the aggregated estimands $\hat \tau_k^{SSDiD}$ are comparable across different horizons $k$, as they average over the same set of cohorts. Second, as we discuss in the next section, treatment effects for some cohorts at long horizons may not be identified, making a uniform maximum horizon a theoretically grounded choice.
\end{remark}

\begin{remark}
An important feature of the Sequential SDiD is its connection to simpler DiD estimators. If we set the regularization parameter $\eta = \infty$, the weight-finding problem reduces to selecting uniform weights (inversely proportional to cohort shares for $\omega$). The resulting estimator becomes equivalent to a sequential DiD estimator. As we show later, this special case is closely related to recent proposals in the event-study literature, particularly the imputation estimator of \cite{borusyak2024revisiting}. In our simulations, we use this version of the estimator (denoted $\hat \tau_k^{DiD}$) as a benchmark.
\end{remark}

\subsection{Inference and validation}
To conduct inference on the estimated treatment effects, we rely on the Bayesian bootstrap \citep{rubin1981bayesian,chamberlain2003nonparametric}. The procedure involves drawing a set of $n$ independent weights $\boldsymbol{\xi} := \{\xi_i\}_{i=1}^n$ from an Exponential distribution, $\xi_i \sim \text{Exp}(1)$. These weights are used to construct a bootstrapped version of the aggregate outcomes:
\begin{equation*}
 Y_{a,t}(\boldsymbol{\xi}) := \frac{\sum_{i:A_i =a}Y_{i,t}\xi_i}{\sum_{i: A_i =a} \xi_i}.
\end{equation*}
We then apply Algorithm \ref{alg:ssdid} to these re-weighted outcomes $Y_{a,t}(\boldsymbol{\xi})$ to produce a bootstrap replicate of our estimator, $\hat \tau_k^{SSDiD}(\mu, \boldsymbol{\xi})$. By repeating this process many times, we can simulate the conditional distribution of the estimator given the data, and use the quantiles of this distribution to form confidence intervals.

A simpler alternative, which we use in our simulations, is to construct normal-approximation confidence intervals. This involves computing the standard deviation of the bootstrap replicates,
\begin{equation*}
\hat \sigma(\hat \tau_k^{SSDiD}(\mu)) := \sqrt{\mathbb{V}_{\boldsymbol{\xi}}\left[\hat \tau_k^{SSDiD}(\mu, \boldsymbol{\xi}) | \text{Data} \right]},
\end{equation*}
and using it to form a standard Wald-type interval:
\begin{equation*}
\hat \tau_k^{SSDiD}(\mu) \pm q_{1-\alpha/2} \cdot \hat \sigma(\hat \tau_k^{SSDiD}(\mu)),
\end{equation*}
where $q_{1-\alpha/2}$ is the $1-\frac{\alpha}{2}$ quantile of the standard normal distribution. While we expect such intervals may be conservative in some settings \citep{hahn2021bootstrap}, we find that they perform reasonably well in our simulations. 

We also adapt our method for placebo validation, analogous to testing for pre-trends in a standard DiD analysis. For a chosen placebo horizon $P > 0$, we define a set of placebo adoption times by shifting the true adoption times backward:
\begin{equation*}
A_i^P := A_i - P.
\end{equation*}
We then construct aggregate outcomes based on these placebo cohorts $A_i^P$ and apply Algorithm \ref{alg:ssdid} to this new dataset, setting the maximum horizon to $K = P-1$. Under the model's identifying assumptions, the resulting placebo estimates, $\hat{\tau}_{a,k}^{SSDiD}$ for $k \in \{0, \dots, P-1\}$, should be centered at zero. Significant deviations from zero would cast doubt on the validity of the model.

\begin{remark}
 This placebo framework can also be adapted to produce valid estimates under alternative assumptions. For instance, if the no-anticipation condition in Assumption \ref{as:potential outcomes} is violated, but an assumption of limited anticipation holds (e.g., units react at most $P$ periods before their official adoption), one can still obtain valid estimates. By using the shifted adoption times $A_i^P $ as the event dates and setting the estimation horizon $K > P-1$, the estimator will target the actual treatment effects for horizons $k \ge P$, as discussed in \cite{callaway2021difference}.
\end{remark}

\section{Theoretical Analysis}\label{sec:theory}
In this section, we establish the theoretical properties of our estimator. Our analysis proceeds in two main steps: we first introduce an infeasible "oracle" OLS estimator that serves as a natural benchmark, and then we show that our Sequential SDiD estimator is asymptotically equivalent to this oracle. This approach has two distinct advantages. First, the oracle estimator we consider is grounded in established empirical practice, making the link between our procedure and what applied researchers might ideally want to do explicit and immediately valuable. Second, the analysis of the oracle uncovers new algorithmic properties of OLS in staggered adoption settings, a result of independent interest. All proofs are collected in the Appendix.

\subsection{Sequential OLS}
We begin our theoretical analysis with an infeasible "oracle" OLS estimator constructed from the aggregate data, $Y_{a,t}$. This oracle knows the factor loadings $\theta_a$ and factors $\psi_t$ and uses them as regressors. Specifically, we consider the solution to the following weighted least squares problem:
\begin{equation}\label{eq:or_ols}
\begin{split}
\{\hat\alpha^{OLS}_a,\hat\beta^{OLS}_t, \hat\phi^{OLS}_t, \hat\nu^{OLS}_a,\hat\tau^{OLS}_{a,k}\}_{a,t,k} \in 
\argmin_{\{\alpha_a,\beta_t, \phi_t, \nu_a,\tau_{a,k}\}} & \sum_{a,t}\pi_a \left(Y_{a,t} - \alpha_a - \beta_t - \theta_a^\top \phi_t - \nu_a^\top\psi_t \right. \\
& \left. - \mathbf{1}\{a \le t\}\tau_{a,t-a} \right)^2.
\end{split}
\end{equation}
The optimization in \eqref{eq:or_ols} treats the factor components $\theta_{a}$ and $\psi_t$ as known regressors to be interacted with unknown cohort-specific and period-specific coefficients, $\nu_a$ and $\phi_t$. This can be viewed as a tractable linearization of the fully nonlinear interactive fixed effects problem. As we will show, the solution to this problem is intimately connected to our proposed Sequential SDiD estimator.

A priori, it is not obvious whether the parameters of interest, $\hat \tau^{OLS}_{a,k}$, are uniquely defined by \eqref{eq:or_ols}. In a standard two-way fixed effects model, the conditions for identification are straightforward, requiring only that a valid comparison group exists. The situation is more challenging here. In principle, it is possible for an effect at a longer horizon, $\hat \tau^{OLS}_{a,l}$ (for $l>k$), to be identified even if the shorter-horizon effect $\hat \tau^{OLS}_{a,k}$ is not. Our next assumption provides a sufficient condition to ensure orderly identification.
\begin{assumption}[Affine Hull]\label{as:rank_cond}
There exist periods $a^{\star}$ and $t^{\star}$ with $a^{\star} \ge t^{\star}$ such that the set of control-group factor loadings $\{\theta_{j}\}_{j> a^{\star}}$ and the set of pre-treatment factors $\{ \psi_{l}\}_{l<t^{\star}}$ both affinely span $\mathbb{R}^{r}$.
\end{assumption}
Assumption \ref{as:rank_cond} has a natural interpretation. For the time factors $\{\psi_t\}$, it requires that factors far enough in the future ($t \ge t^{\star}$) can be perfectly predicted by an affine combination of past factors. The requirement for the unit-specific loadings $\{\theta_a\}$ is similar: the loadings of early-adopting cohorts can be represented as an affine combination of the loadings of sufficiently late-adopting cohorts. 

This assumption allows us to state our first main result, which is not only crucial for our subsequent analysis but also holds independent interest.
\begin{prop}\label{pr:OLS_comp}
    Suppose Assumption \ref{as:rank_cond} holds. Then for any cohort $a$ and horizon $k$ such that $t^{\star} \le a+k \le a^{\star}$, the OLS estimator $\hat \tau_{a,k}^{OLS}$ is uniquely defined and can be computed using Algorithm \ref{alg:ols}.
\end{prop}
Proposition \ref{pr:OLS_comp} is important because it provides an explicit, sequential algorithm for computing the oracle OLS estimates. The structure of Algorithm \ref{alg:ols} deliberately mirrors that of our main proposal, Sequential SDiD (Algorithm \ref{alg:ssdid}). For each $(a,k)$, it constructs unit- and time-specific weights, $\tilde \omega^{(a)}$ and $\tilde \lambda^{(a,k)}$, and uses them to form a weighted DiD estimator. It then concludes by imputing the counterfactual outcome and proceeds sequentially.

\RestyleAlgo{boxruled}
\LinesNumbered
\begin{algorithm}[t!]
 \KwData{Aggregated data $\{Y_{a,t}, \theta_a, \psi_t, \pi_a\}$, parameters $a^{\star}$, $t^{\star}$}
 \KwResult{OLS estimates $\{\hat \tau_{a,k}^{OLS}\}$ for $t^{\star} \le a+k \le a^{\star}$}
 \For{$k \in \{0, \dots, a^{\star} - t^{\star}\}$}{
  \For{ $a \in \{t^{\star},\dots, a^{\star} - k\}$}{
   Construct the oracle weights:
    
    \begin{equation*}
    \begin{aligned}
     \tilde\omega^{(a)} := \argmin_{\omega} \left\{\sum_{j > a} \frac{\omega_j^2}{\pi_j}\right\} \quad
     &\text{s.t.} && \sum_{j > a} \omega_j = 1, \quad \sum_{j >a} \theta_j \omega_j = \theta_{a}; \\
     \tilde\lambda^{(a,k)} := \argmin_{\lambda} \left\{\sum_{l < a+ k} \lambda_l^2\right\} \quad
     &\text{s.t.} && \sum_{l <a+k} \lambda_l = 1, \quad \sum_{l < a+ k} \lambda_l \psi_l = \psi_{a+k};
    \end{aligned}
    \end{equation*}
   \nl Construct the estimator:
    \begin{equation*}
         \hat \tau_{a,k}^{OLS} := \left(Y_{a,a+k} - \sum_{j>a} \tilde\omega_j^{(a)}Y_{j,a+k}\right) -  \sum_{l <a+k}\tilde\lambda_l^{(a,k)}\left(Y_{a,l} - \sum_{j>a}\tilde\omega_j^{(a)}Y_{j,l}\right);
    \end{equation*}
   \nl Impute the counterfactual for the next iteration:
    \begin{equation*}
      Y_{a,a+k} := Y_{a,a +k} - \hat \tau_{a,k}^{OLS};
    \end{equation*}
  }
 }
\caption{Sequential OLS}\label{alg:ols}
\end{algorithm}

The sequential representation of $\hat \tau_{a,k}^{OLS}$ in Algorithm \ref{alg:ols} is key to establishing the statistical connection between the oracle OLS and the feasible Sequential SDiD estimators in the next section. However, this result is also interesting in its own right as it reveals the underlying mechanics of OLS in this setting. For instance, if we consider a standard two-way model (i.e., $\theta_a \equiv \psi_l \equiv 0$), Assumption \ref{as:rank_cond} holds trivially. In this case, Algorithm \ref{alg:ols} provides a sequential implementation of the imputation estimator proposed by \cite{borusyak2024revisiting}. The weights $\tilde{\lambda}$ become uniform, and the weights $\tilde{\omega}$ are inversely proportional to cohort size, reducing the procedure to a sequence of standard DiD estimators applied to imputed data.

The sequential nature of Algorithm \ref{alg:ols} also has practical implications, showing that $\hat \tau_{a,k}^{OLS}$ can, in principle, be computed "online" using only information available up to period $a+k$. While less critical for the small-scale problems we consider, this property is valuable in large-scale industrial applications. Finally, this representation opens several paths for generalizations of OLS, with our Sequential SDiD estimator being one such path. Other routes could include adding further regularization to the weights (e.g., a simplex constraint) or restricting the information set to allow for weaker exogeneity assumptions.

\begin{remark}
The representation we derive in Proposition \ref{pr:OLS_comp} is not unique. For a model without interactive fixed effects, \cite{aguilar2023estimation} derives a non-sequential representation of the OLS estimator. For the reasons described above, we believe our sequential representation offers multiple analytical and practical advantages.
\end{remark}

\subsection{Sequential SDiD vs. Sequential OLS}
In this section, we formally connect our feasible Sequential SDiD estimator (Algorithm \ref{alg:ssdid}) to the infeasible oracle OLS estimator (Algorithm \ref{alg:ols}). As the procedural similarities suggest, the two are closely related. To establish this statistical relationship, however, we require some mild restrictions on the data-generating process.

We begin by imposing a regularity condition on the aggregate errors, $\boldsymbol{\epsilon}_a := (\epsilon_{a,1},\dots, \epsilon_{a,T})$.
\begin{assumption}[Regularity of Aggregate Errors]\label{as:var_behavior}
For all $a \in \mathcal{A}$, the scaled variance-covariance matrix of the aggregate errors converges to a finite and non-degenerate limit: $n\mathbb{V}[\boldsymbol{\epsilon}_{a}]\rightarrow \Sigma_a$.
\end{assumption}
Assumption \ref{as:var_behavior} is a mild condition that ensures the cohort-level noise diminishes at a standard $\frac{1}{\sqrt{n}}$ rate. It implicitly requires that the share of each adoption cohort, $n_a/n$, is non-vanishing as $n \to \infty$, which is central to our large-$n$, fixed-$T$ asymptotic framework.

To state our main result, we introduce additional notation. For each relevant cohort-horizon pair $(a,k)$, we define a matrix $L^{(a,k)}$ that captures the demeaned interactive fixed effects:
\begin{equation*}
(L^{(a,k)})_{j,l} := \left(\theta_j - \overline{\theta}^{(a)}\right)^\top \left(\psi_{l} - \overline{\psi}^{(k)}\right),
\end{equation*}
for cohorts $j >a$ and periods $l <a+k$. Here, $\overline{\theta}^{(a)}$ and $\overline{\psi}^{(k)}$ are the averages of the respective factor components over the relevant control units and pre-treatment periods. Assumption \ref{as:rank_cond} guarantees that for the $(a,k)$ pairs we consider, this matrix has full rank $r$. We use $\tilde \sigma_{(a,k)}$ to denote the smallest non-zero singular value of $L^{(a,k)}$.

We are now ready to state the main theorem, which establishes the asymptotic equivalence between the Sequential SDiD and the oracle OLS estimator.
\begin{theorem}\label{th:as_connection}
Suppose the following conditions hold:
\begin{enumerate}
\item The model satisfies Assumptions \ref{as:potential outcomes}, \ref{as:factor}, \ref{as:rank_cond}, and \ref{as:var_behavior}.
\item The oracle OLS estimator has a bounded variance: for all relevant $(a,k)$, we have
\begin{align*}
 n\mathbb{V}[\hat \tau_{a,k}^{OLS}|\boldsymbol{\gamma},\{A_i\}_{i=1}^n]\lesssim 1
\end{align*}
\item The factors are not too weak: the minimal singular value of the demeaned factor matrix vanishes slower than the noise level, i.e., $\tilde \sigma_{a,k} \gg_p n^{-1/2}$.
\item The regularization parameter $\eta$ is chosen within the appropriate range: $(n^{-\frac12} \tilde \sigma^3_{a,k})^{\frac14}\gg \eta \gg n^{-\frac12}$.
\end{enumerate}
Then, the Sequential SDiD estimator is asymptotically equivalent to the oracle OLS estimator:
\begin{equation*}
\hat \tau_{a,k}^{SSDiD} = \hat \tau_{a,k}^{OLS} + o_p\left(\frac{1}{\sqrt{n}}\right).
\end{equation*}
\end{theorem}
Theorem \ref{th:as_connection} provides the key theoretical justification for our method. The conditions are relatively mild. We require the oracle OLS estimator to be well-behaved (Condition 2), which is natural for a benchmark. The key restriction is on the strength of the factors (Condition 3). This condition requires the identifying variation from the interactive fixed effects to be stronger than the statistical noise, but it allows the factors to be "weak" in the sense that their explanatory power can vanish as $n \to \infty$. Finally, we require the regularization parameter to be chosen appropriately (Condition 4), balancing the bias from regularization with the variance from overfitting.

As long as these conditions hold, our feasible Sequential SDiD estimator mimics the infeasible oracle. This result is particularly relevant because the "weak factor" case is common in applications. While standard two-way fixed effects often explain a large share of the variance in aggregated data, the incremental contribution of interactive fixed effects can be small, yet still larger than the noise and thus important for valid inference \citep{arkhangelsky2021synthetic}. Our result, which allows the weakest factor to be only marginally stronger than the noise level ($n^{-1/2}$), is therefore especially appealing for empirical work. This contrasts with estimators designed for "strong" factors, which are typically assumed in the large panel data literature (e.g., \citealp{bai2009panel}), and avoids the complexities of bias-aware inference required when factors are "at the noise level" \citep{armstrong2022robust}.

A key practical implication of Theorem \ref{th:as_connection} is that it provides a direct justification for using standard inference procedures, as formalized in the following corollary.
\begin{corollary}\label{cor:bootstrap}
Under the conditions of Theorem \ref{th:as_connection}, the Bayesian bootstrap procedure is asymptotically valid for constructing confidence intervals for $\tau_{a,k}$ and its averages. 
\end{corollary}

\begin{remark}[Choosing the Regularization Parameter]\label{rem:eta_choice}
While the bounds on $\eta$ in Condition 4 of Theorem \ref{th:as_connection} appear complex, their role is to ensure the regularization is strong enough to suppress statistical noise (the $\eta \gg n^{-1/2}$ condition) but not so strong that it creates significant bias by oversmoothing. The condition allows for a wide range of valid choices. Importantly, simple rules of thumb, such as setting $\eta^2$ to be proportional to the variance of noise scaled by the sample size (e.g., $\eta^2 \propto \frac{\sigma^2}{n^{0.9}}$ will satisfy these theoretical requirements. We use such a practical rule in our application in Section~\ref{sec:exp}.
\end{remark}

\begin{remark}
We conjecture that an analog of Theorem \ref{th:as_connection} also holds in the "very weak" factor regime, where the largest singular value of $L^{(a,k)}$ also vanishes. In that case, the relevant oracle would likely be one that ignores factors below the noise level. For the extreme case where all factors are zero (i.e., the standard two-way model holds), it is straightforward to show that our estimator is asymptotically equivalent to the standard OLS estimator. A full investigation of the transition between these regimes, however, would require a more nuanced analysis that we leave for future work.
\end{remark}

\subsection{Efficiency}
The asymptotic equivalence established in Theorem \ref{th:as_connection} allows us to analyze the efficiency of our Sequential SDiD estimator by studying the well-understood properties of the oracle OLS estimator. Because $\hat{\tau}_{a,k}^{SSDiD}$ is first-order equivalent to $\hat{\tau}_{a,k}^{OLS}$, it inherits its asymptotic efficiency properties.

The most direct efficiency result stems from the Gauss-Markov theorem. If the aggregate errors are homoskedastic and serially uncorrelated for each cohort (i.e., $\mathbb{V}[\boldsymbol{\epsilon}_a|n_a] = \frac{\sigma^2}{n\pi_a}\mathcal{I}_{T}$ for some $\sigma^2 > 0$), then the oracle estimator in \eqref{eq:or_ols} is the Best Linear Unbiased Estimator (BLUE). This guarantee is analogous to the one established by \cite{borusyak2024revisiting} for their imputation procedure under similar conditions. Our results therefore extend this type of efficiency guarantee to the Sequential SDiD estimator. To our knowledge, this provides the first formal efficiency result for an SDiD-type estimator and, more broadly, for any synthetic control-type procedure.

The connection to OLS also clarifies the limits of our estimator's efficiency. While OLS is BLUE under spherical errors, it is not, in general, the most efficient estimator in the presence of heteroskedasticity or serial correlation; Generalized Least Squares (GLS) would be preferred. While we do not formally establish this, we anticipate that under the i.i.d model (Example 1 in Section~\ref{sec:method}) the relevant limit experiment for estimating $\tau_{a,k}$ is a normal model with a mean structure given by \eqref{eq:or_ols} but with a non-spherical error covariance. In such cases, the semiparametric efficiency bound would be attained by a GLS-type estimator. This implies that while our proposed estimator is efficient within a class that mirrors common empirical practice, we do not expect it to be semiparametrically efficient without further assumptions or modifications.

\section{Covariates}\label{sec:cov}
Our analysis thus far has abstracted from covariates. In practice, however, researchers usually observe both time-invariant characteristics, which we denote by $X_i$, and time-varying characteristics, $Z_{i,t}$. In this section, we discuss how to incorporate such information into our framework.

Our focus is on time-invariant covariates $X_i$ (assumed to be discrete), for two reasons. First, from a practical standpoint, interacting time fixed effects with time-invariant characteristics is a powerful and common way to control for unobserved heterogeneity. Time-varying controls $Z_{i,t}$, in our experience, often have limited additional predictive power in many applications.\footnote{This is the case in the empirical example we analyze in Section \ref{sec:exp}.} Second, there is a theoretical challenge with time-varying covariates: if $Z_{i,t}$ has a substantial, dynamic causal effect on the outcome, it essentially behaves like an additional treatment variable. Properly analyzing such multi-treatment settings requires strong assumptions about effect homogeneity and is beyond the scope of this paper (see, e.g., \citealp{de2023two}).

We therefore focus on three distinct strategies for incorporating discrete, time-invariant covariates $X_i$ into our framework, each suited to a different empirical context.

\paragraph{Strategy 1: Full Stratification.}
The most flexible approach is to allow all model parameters to vary with the value of the covariate, $x \in \mathcal{X}$. This corresponds to a fully-interacted model:
\begin{equation}\label{eq:mod_cov}
    Y_{i,t} = \alpha_{i} +\beta_t(X_i) + \theta_i^{\top}\psi_t(X_i) + \sum_{k\ge 0}\tau_{i,a,k}(X_i)\mathbf{1}\{A_i = a,k = t-A_i\} + \epsilon_{i,t}.
\end{equation}
This approach is equivalent to stratifying the data by each value of $x$ and applying our main algorithm separately to each stratum. While conceptually straightforward, this is often impractical. If the set of covariates $\mathcal{X}$ is large, the resulting subsamples for each $(a,x)$ pair may be too small for our asymptotic arguments to hold, leading to noisy and unstable estimates. We therefore do not recommend this strategy unless every stratum is large.

\paragraph{Strategy 2: A Practical Hybrid Model (Recommended Approach).}
Our recommended strategy for most applications is to adopt a more parsimonious, hybrid model. We allow the additive unit- and time-specific fixed effects to depend on $X_i$, but assume the multiplicative factors $\psi_t$ are common across all units:
\begin{equation}\label{eq:mod_cov_sim}
     Y_{i,t} = \alpha_{i} +\beta_t(X_i) + \theta_i^{\top}\psi_t+ \sum_{k\ge 0}\tau_{i,a,k}\mathbf{1}\{A_i = a,k = t-A_i\} + \epsilon_{i,t},
\end{equation}
where we assume $\mathbb{E}[\epsilon_{i,t}|A_i, X_i] = 0$. This specification is more restrictive than the fully stratified model in \eqref{eq:mod_cov}, but it remains more general than the standard conditional parallel trends assumption made in much of the literature (e.g., \citealp{abadie2005semiparametric,sant2020doubly}). It is also particularly natural if we view $\psi_t$ as representing pervasive aggregate shocks, as in Example 2 of Section \ref{subsec:setup}.

The key advantage of model \eqref{eq:mod_cov_sim} is that it behaves well under aggregation. A researcher can first compute cohort- and covariate-specific averages, $Y_{a,t}(x)$, and then average these over $x$ to produce a single aggregate series, $Y_{a,t}$, for each cohort $a$. The resulting series will have exactly the same structure as in our main model \eqref{eq:agg_eq}, allowing for the direct application of Algorithm \ref{alg:ssdid}. This approach is robust to arbitrary heterogeneity in the treatment effects across covariate groups and is our recommended strategy in the common case where adoption times vary within covariate groups (i.e., $A_i$ is not a deterministic function of $X_i$).

A special consideration arises for the never-treated cohort ($a=\infty$). Since we do not estimate treatment effects for this group, the outcomes $Y_{\infty, t}(x)$ for different $x$ can be aggregated using flexible, data-driven weights. In practice, if the number of never-treated units is large, we suggest including each $Y_{\infty, t}(x)$ as a separate control series in the synthetic control algorithm.

\paragraph{Strategy 3: The Case of Group-Level Treatment Assignment.}
Finally, we consider a different approach for the specific but important case where adoption time is a deterministic function of the covariates, i.e., $A_i = A(X_i)$. This is common in applications where $X_i$ represents a geographic unit like a state or county. 
In this setting, the model with conditional parallel trends is too flexible to allow for identification, and researchers often rely on the stronger assumption of unconditional parallel trends. 
As an attractive alternative, our framework allows one to strengthen the exogeneity assumption using $X_i$ without including it directly in the functional form for the outcome:
\begin{equation}\label{eq:mod_cov_inst}
     Y_{i,t} = \alpha_{i} +\beta_t + \theta_i^{\top}\psi_t+ \sum_{k\ge 0}\tau_{i,a,k}\mathbf{1}\{A_i = a,k = t-A_i\} + \epsilon_{i,t}, \quad \text{with } \mathbb{E}[\epsilon_{i,t}|X_i] = 0.
\end{equation}
In this case, rather than averaging outcomes within adoption cohorts, it is natural to average them within the groups defined by $X_i$. We illustrate this strategy in one of the simulations in the next section.

\section{Empirical illustration and simulations}\label{sec:exp}

\subsection{Empirical Example}
We first apply our method to reevaluate the findings of \cite{bailey2015war}, who study the effect of Community Health Centers (CHCs) on mortality rates. The original dataset consists of county-year level observations from 1959-1988. We follow the authors' main sample construction choices and exclude the three most populous counties, which were all treated in the same year, to ensure a more balanced design. The main outcome, $Y_{i,t}$, is the adjusted mortality rate, and the treatment, $W_{i,t}$, is the presence of a CHC in a given county and year.

To implement our procedure, we first aggregate the data. For treated counties, we define cohorts based on their CHC adoption date ($A_i$). For never-treated counties, we create five distinct control cohorts based on their 1960 urban population percentage, a key baseline covariate. This follows the logic of our Strategy 2 from Section \ref{sec:cov}. All cohort-level outcomes, $Y_{a,t}$, are calculated as weighted averages, using 1960 county populations as weights.

We then compute our Sequential SDiD estimates using Algorithm \ref{alg:ssdid}, setting the regularization parameter $\eta^2 = \hat{\sigma}^2/n^{0.9}$, where $\hat{\sigma}^2$ is a preliminary estimate of the error variance from a standard two-way fixed effects model. For comparison, we also compute standard DiD estimates by setting $\eta = \infty$. Standard errors are produced via the Bayesian bootstrap with $1000$ replications.

\begin{figure}[t!]
    \begin{center}
    \caption{Effect of Community Health Centers on Mortality}
    \label{fig:reps}
    \includegraphics[scale = 0.5]{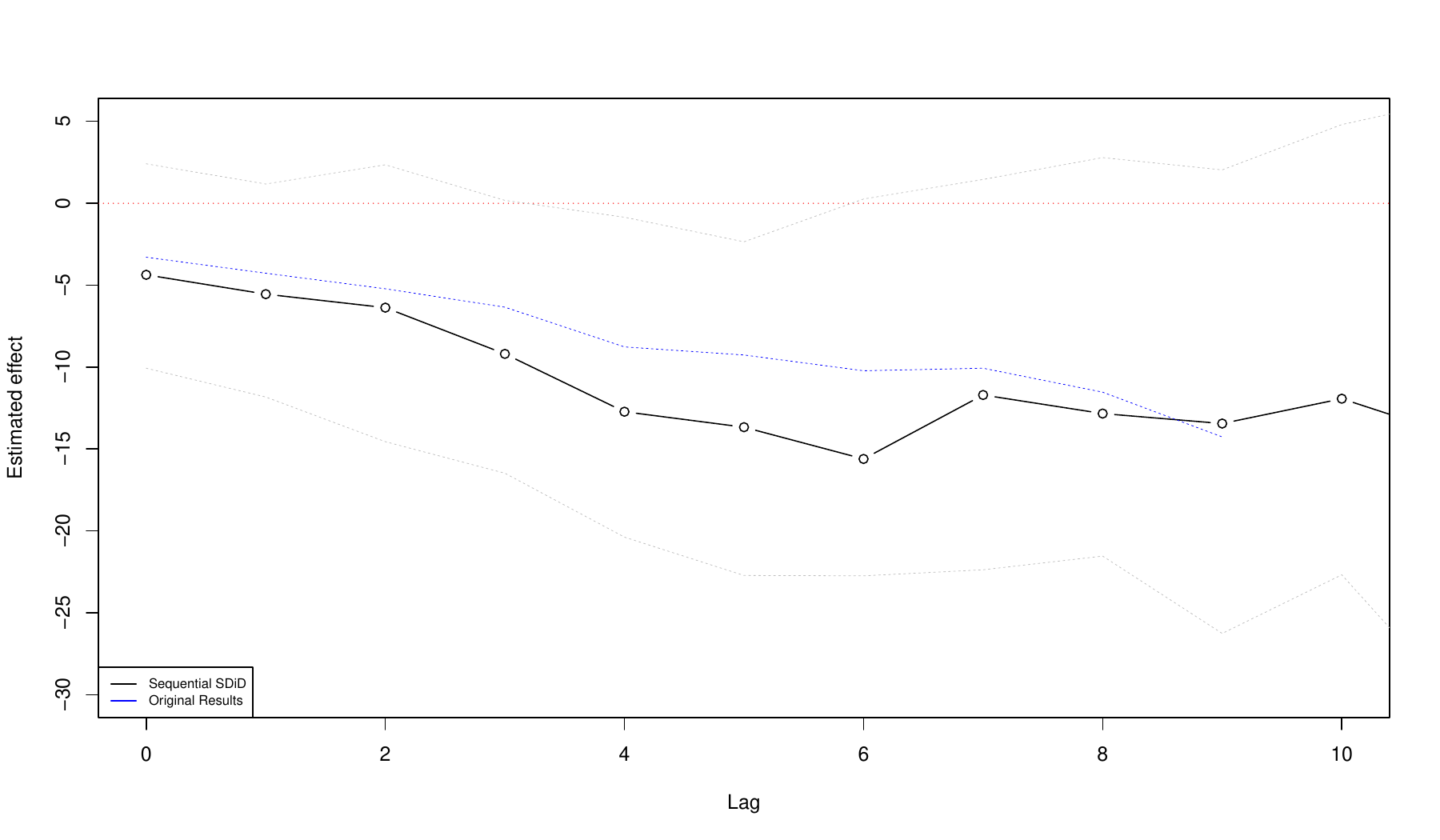}
    \end{center}
 \renewcommand{\baselinestretch}{0.7}
\footnotesize{\textit{Notes}: The figure plots the estimated average treatment effect on the treated ($\hat{\tau}_k$) for $k$ years after CHC adoption. The "Original Results" line corresponds to a sequential DiD estimator ($\eta = \infty$). The "Sequential SDiD" line is our main estimator. The shaded region represents the 95\% confidence interval for the Sequential SDiD estimates, constructed via Bayesian bootstrap with 1000 replications.}
\end{figure}

Figure \ref{fig:reps} plots the resulting estimates. As shown, the Sequential SDiD estimates closely track the standard DiD estimates. This is expected, as the original study's pre-trends analysis suggests that the two-way fixed effects model is a good approximation for this application. We view this as a successful "proof of concept," demonstrating that our method produces sensible results in settings where simpler methods are believed to perform well.

\subsection{Simulation Experiments}
\subsubsection{Experiment 1: Robustness to Unobserved Factors}
We now use a simulation study to assess our estimator's performance when the parallel trends assumption fails due to unobserved interactive fixed effects. Our simulation design is based on the \cite{bailey2015war} data. First, we use the matrix completion method of \cite{athey2021matrix} on the original data to create a complete panel of potential outcomes under no treatment, $Y_{i,t}(\infty)$. From this completed panel, we estimate and extract the interactive fixed effects component, which we find to have a rank of 5.

We then generate simulated data where the strength of these interactive fixed effects varies. We define a "signal" strength parameter, where a 0\% signal corresponds to a standard two-way fixed effects model (no interactive effects), and an 80\% signal corresponds to a model where the variance of the interactive fixed effects component is four times larger than the variance of the idiosyncratic noise. To ensure our cohorts are large enough for the asymptotics to be relevant, we expand the sample size by replicating each unit's data (including its treatment status and factor loadings) four times.

For each signal level, we run 1000 simulations. In each simulation, we apply both our Sequential SDiD estimator and the standard DiD estimator, and we compute a $t$-statistic for the estimated treatment effect using 100 bootstrap replications to calculate the standard error.

\begin{figure}[t!]
    \begin{center}
    \caption{Distribution of $t$-statistics for the Contemporaneous Effect ($\tau_0$)}
    \label{fig:t_stat}
    \includegraphics[scale=0.5]{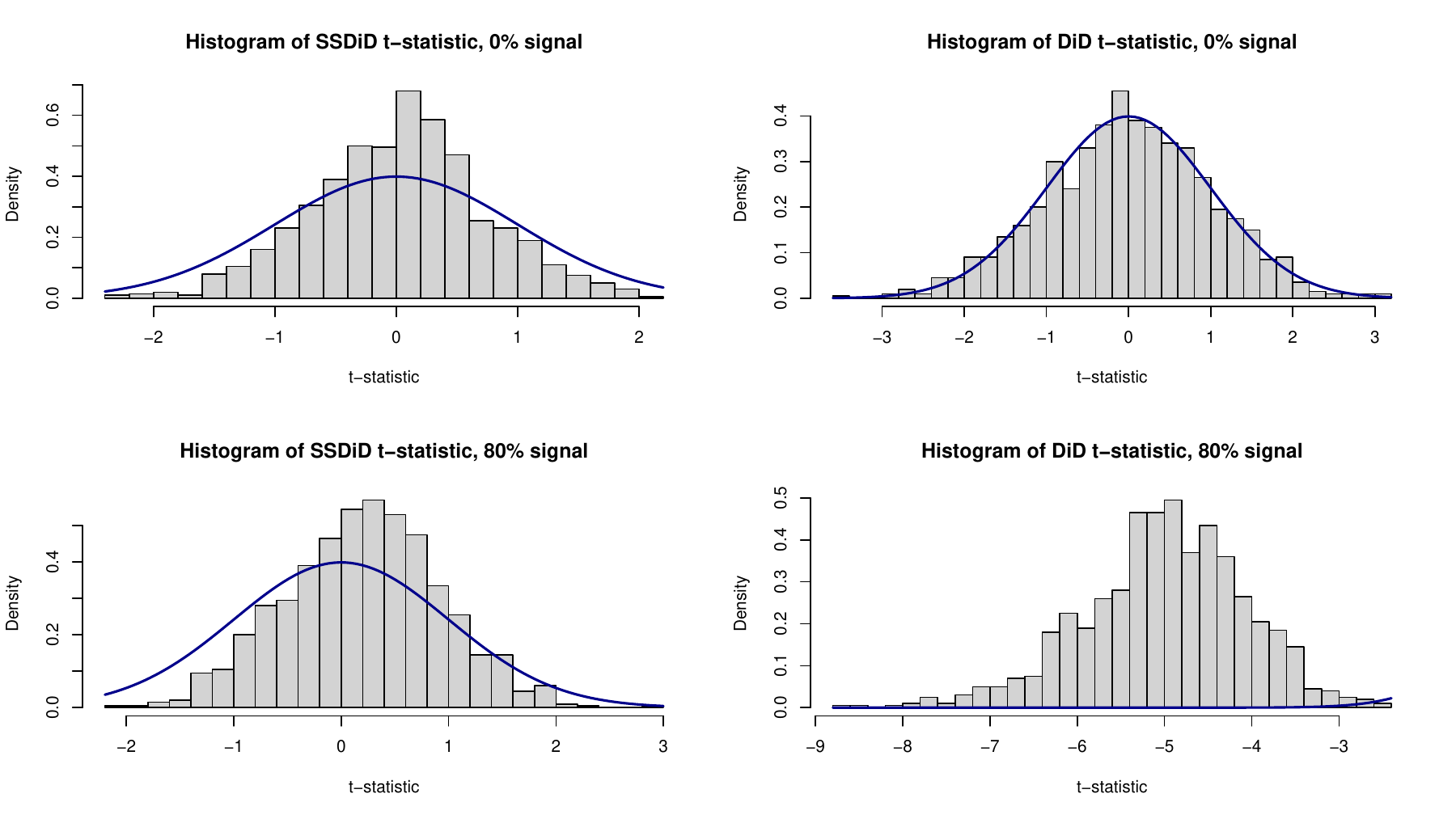}
    \end{center}
\renewcommand{\baselinestretch}{0.7}
    \footnotesize{\textit{Notes}: The histograms show the distribution of t-statistics over 1000 simulations. The top row corresponds to a 0\% signal from interactive fixed effects (i.e., the DiD model is correct). The bottom row corresponds to an 80\% signal. The left column shows results for our Sequential SDiD estimator; the right column shows results for the standard DiD estimator. The blue line is the standard normal density.}
\end{figure}

Figure \ref{fig:t_stat} displays the distribution of these t-statistics for the contemporaneous treatment effect, $\tau_0$. In the 0\% signal case (top row), where the DiD model is correctly specified, both estimators perform well, though the t-statistics for SSDiD are slightly more concentrated, suggesting our bootstrap standard errors are modestly conservative. In the 80\% signal case (bottom row), the standard DiD estimator is severely biased; its t-statistics are centered far below zero, implying its confidence intervals would have near-zero coverage. In sharp contrast, the t-statistics for our Sequential SDiD estimator remain centered at zero, demonstrating that it provides valid inference even in the presence of strong confounding factors.

\begin{figure}[t!]
    \begin{center}
    \caption{Distribution of $t$-statistics for the Lagged Effect ($\tau_4$)}
    \label{fig:t_stat_5}
    \includegraphics[scale=0.5]{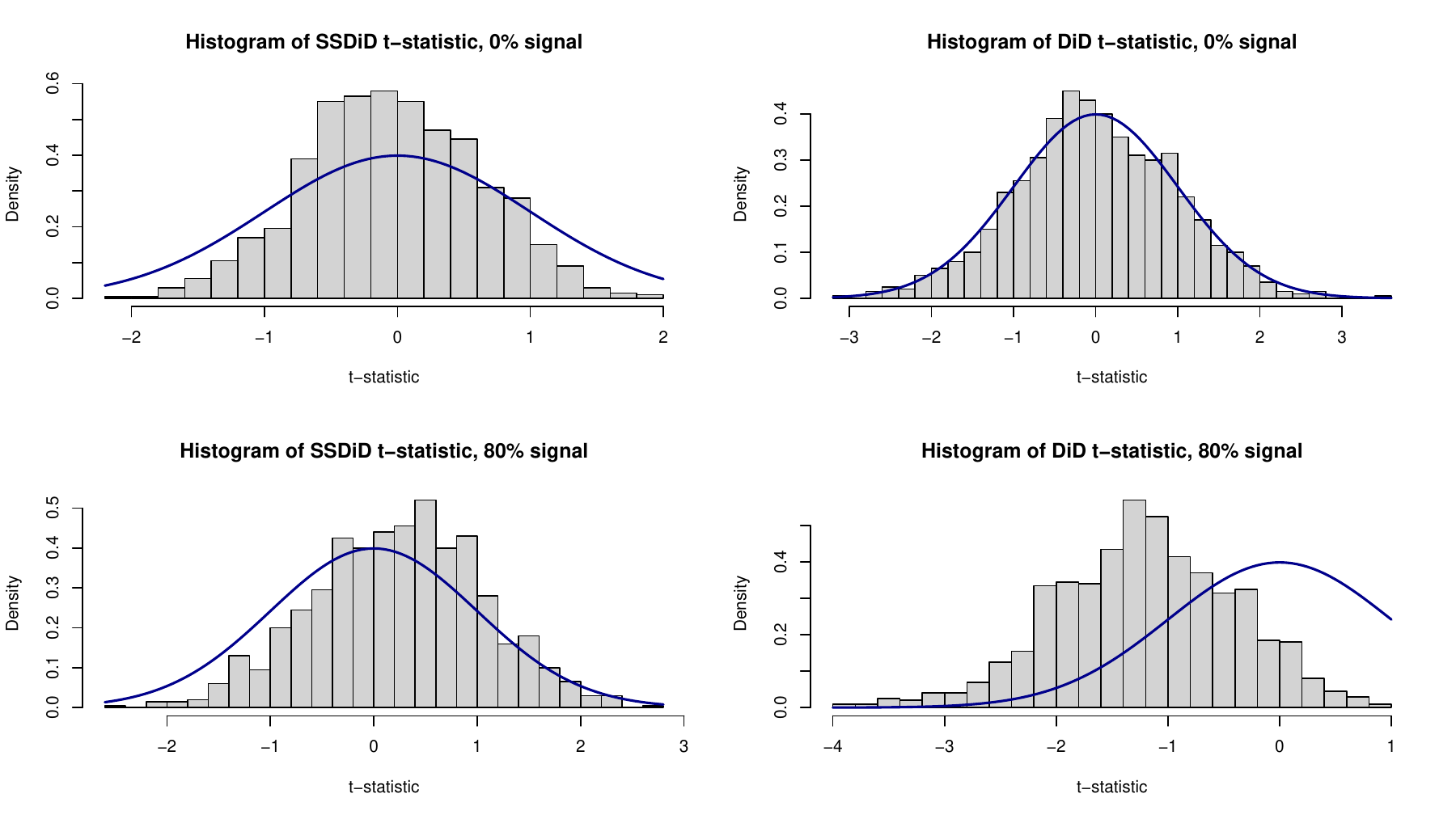}
\end{center}
\renewcommand{\baselinestretch}{0.7}
\footnotesize{\textit{Notes}: The histograms show the distribution of t-statistics for the estimated effect at $k=4$ over 1000 simulations. The top row corresponds to a 0\% signal from interactive fixed effects. The bottom row corresponds to an 80\% signal. The left column shows results for our Sequential SDiD estimator; the right column shows results for the standard DiD estimator. The blue line is the standard normal density.}
\end{figure}

Figure \ref{fig:t_stat_5} repeats this exercise for the treatment effect four years after adoption, $\tau_4$. Again, our estimator performs well in both scenarios. The standard DiD estimator remains biased in the 80\% signal case, although its bias is less extreme. This is likely because the estimation of longer-run effects is inherently noisier, making the bias a smaller component of the overall estimation error. Nonetheless, only the Sequential SDiD estimator provides reliable inference across all scenarios.

\subsubsection{Experiment 2: Calibrated State-Level Panel}

In our second experiment, we assess the performance of our estimator in a setting calibrated to real-world aggregated panel data where interactive fixed effects are empirically important. Inspired by the work of \citet{Bertrand2004did}, the simulation design uses a state-by-year panel of average log wages for women, constructed from the March Current Population Survey (CPS). The panel consists of 50 states over 40 years. Crucially, this design simulates data at the aggregate (state) level, mirroring many empirical applications where researchers do not have access to the underlying microdata. Following the simulation design of \citet{arkhangelsky2021synthetic}, we decompose the observed log-wage panel to estimate its structural components to use for the data-generating process. We find that a standard two-way fixed effects model explains about 94\% of the variation, and an interactive fixed effects component with a rank of $r=4$ explains an additional 5\%. For the simulation, we treat these estimated components as the true, fixed parameters and generate new data by drawing random shocks from the fitted AR(2) process that models the remaining idiosyncratic error.

\begin{figure}[t!]
    \begin{center}
            \includegraphics[width=0.4\linewidth]{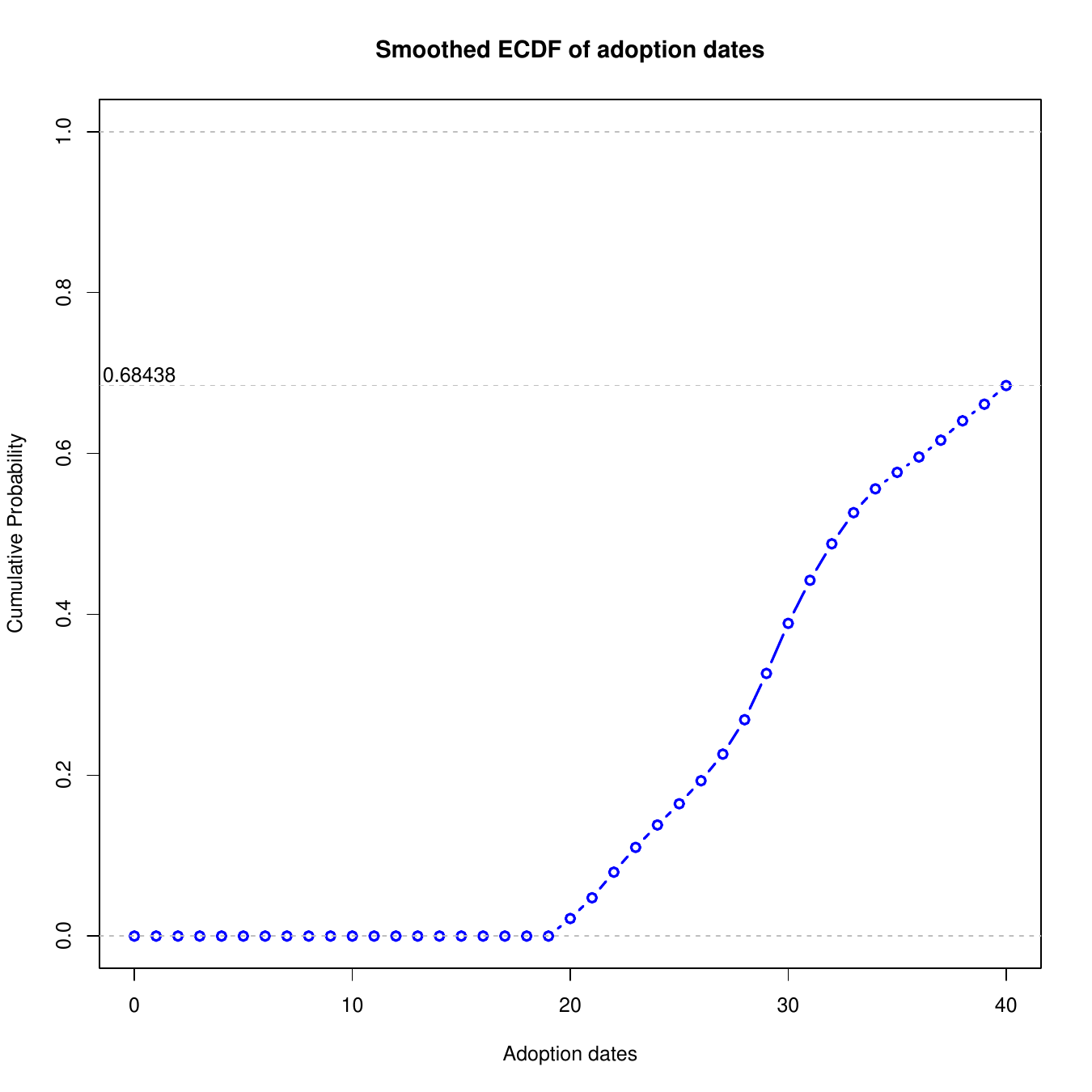}
    \caption{Smoothed Empirical CDF of Simulated Adoption Dates}
    \label{fig:adoption_cdf}
    \end{center}
    \renewcommand{\baselinestretch}{0.7}
\footnotesize{\textit{Notes}: The figure shows the cumulative distribution of adoption dates assigned in the simulation. The assignment mechanism is designed to correlate adoption timing with the underlying interactive fixed effects.}
\end{figure}

A key feature of this experiment is that we induce a correlation between treatment timing and the interactive fixed effects, creating a direct violation of the parallel trends assumption required by standard DiD estimators. We implement a two-stage procedure to assign adoption dates. First, we designate states as "ever-treated" with a probability that depends on the state's factor loading on the first principal component of the interactive-effects model. Second, for these ever-treated states, we assign an adoption date drawn from a normal distribution. The mean of this distribution is also a function of the state's factor loading, ensuring that states with different unobserved trends are systematically treated at different times. The adoption period begins at $t=20$, and a subset of states is always left untreated. Figure~\ref{fig:adoption_cdf} plots the resulting empirical cumulative distribution function of the simulated adoption dates.

\begin{figure}[t!]
\caption{Distribution of $t$-statistics}
\label{fig:state_t_stat}
    \centering

    \begin{minipage}{0.48\textwidth}
        \centering
        \includegraphics[width=\linewidth]{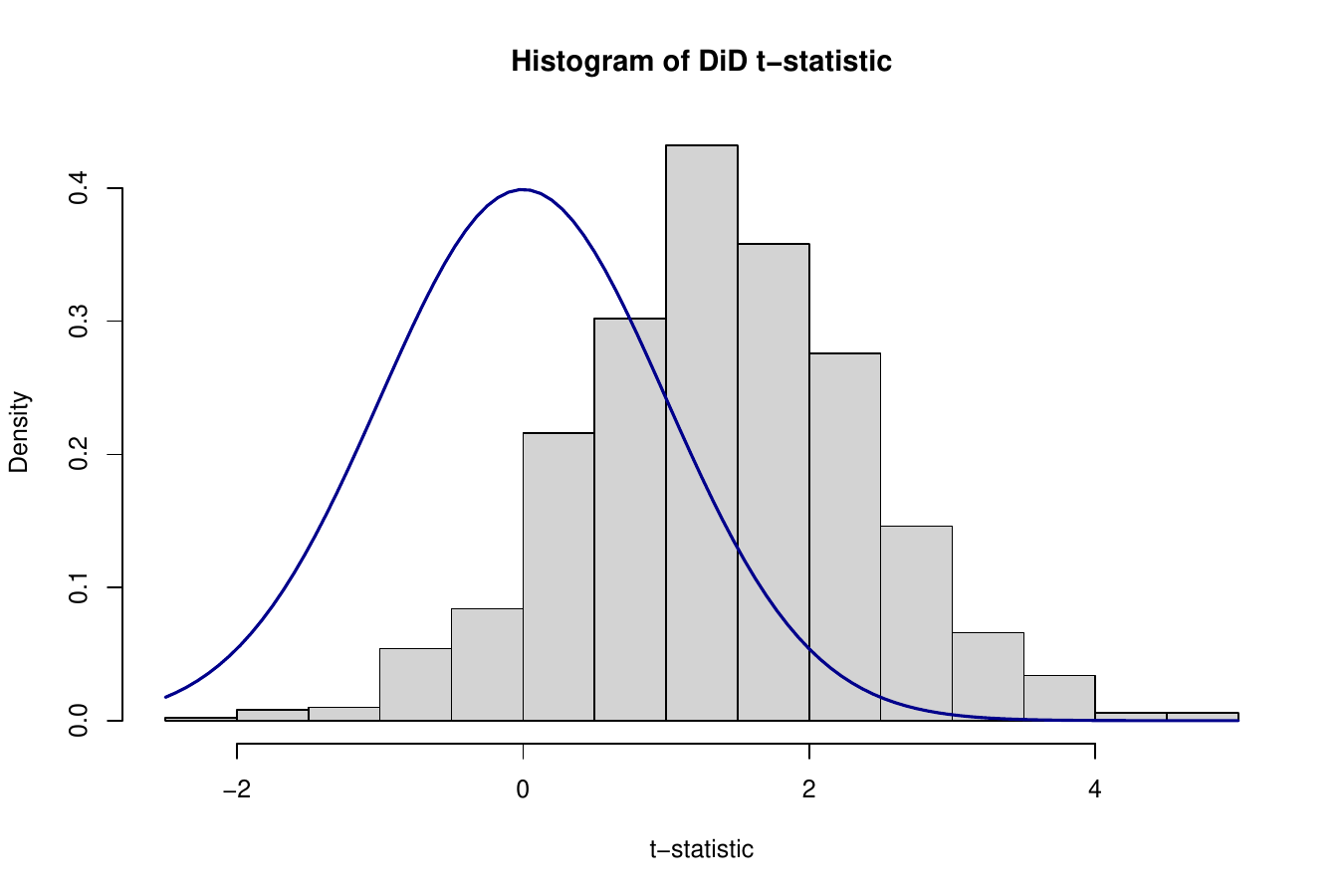}
        \par\vspace{0.5em}
        \small (a) DiD, Contemporaneous Effect ($k=0$)
    \end{minipage}
    \hfill
    \begin{minipage}{0.48\textwidth}
        \centering
        \includegraphics[width=\linewidth]{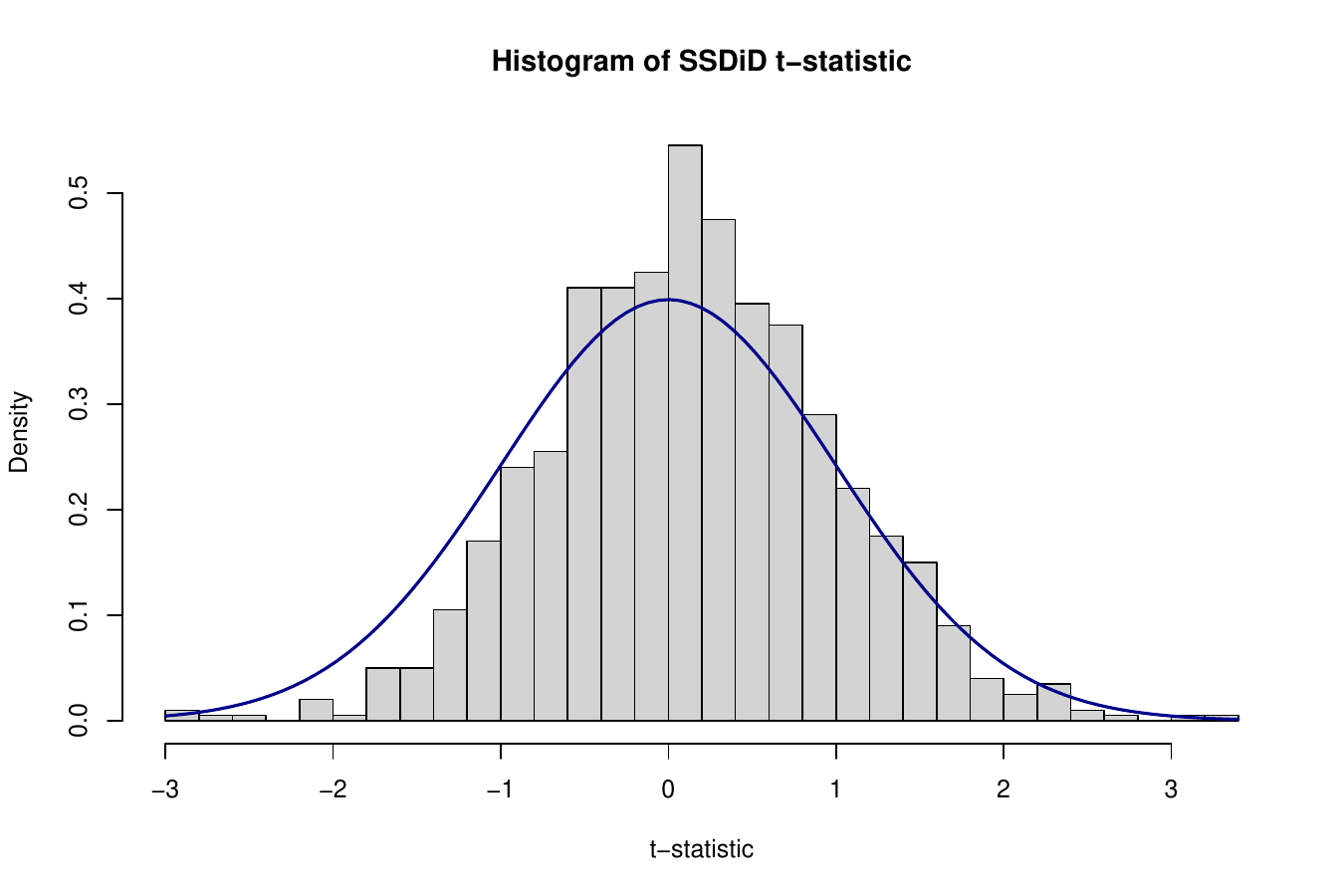}
        \par\vspace{0.5em}
        \small (b) Sequential SDiD, Contemporaneous Effect ($k=0$)
    \end{minipage}

    \vspace{1.5em} 

    \begin{minipage}{0.48\textwidth}
        \centering
        \includegraphics[width=\linewidth]{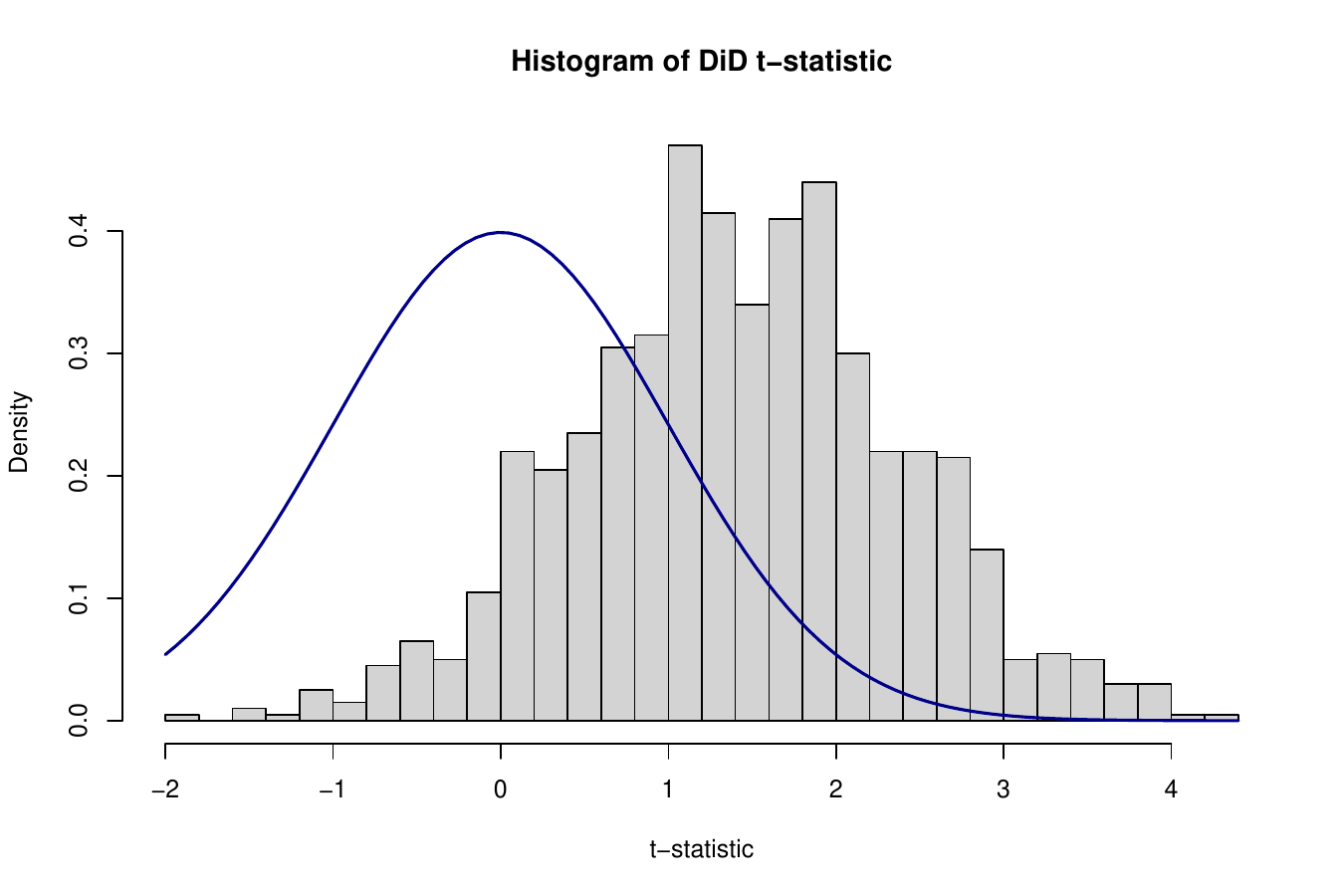}
        \par\vspace{0.5em}
        \small (c) DiD, Fifth-Period Effect ($k=4$)
    \end{minipage}
    \hfill
    \begin{minipage}{0.48\textwidth}
        \centering
        \includegraphics[width=\linewidth]{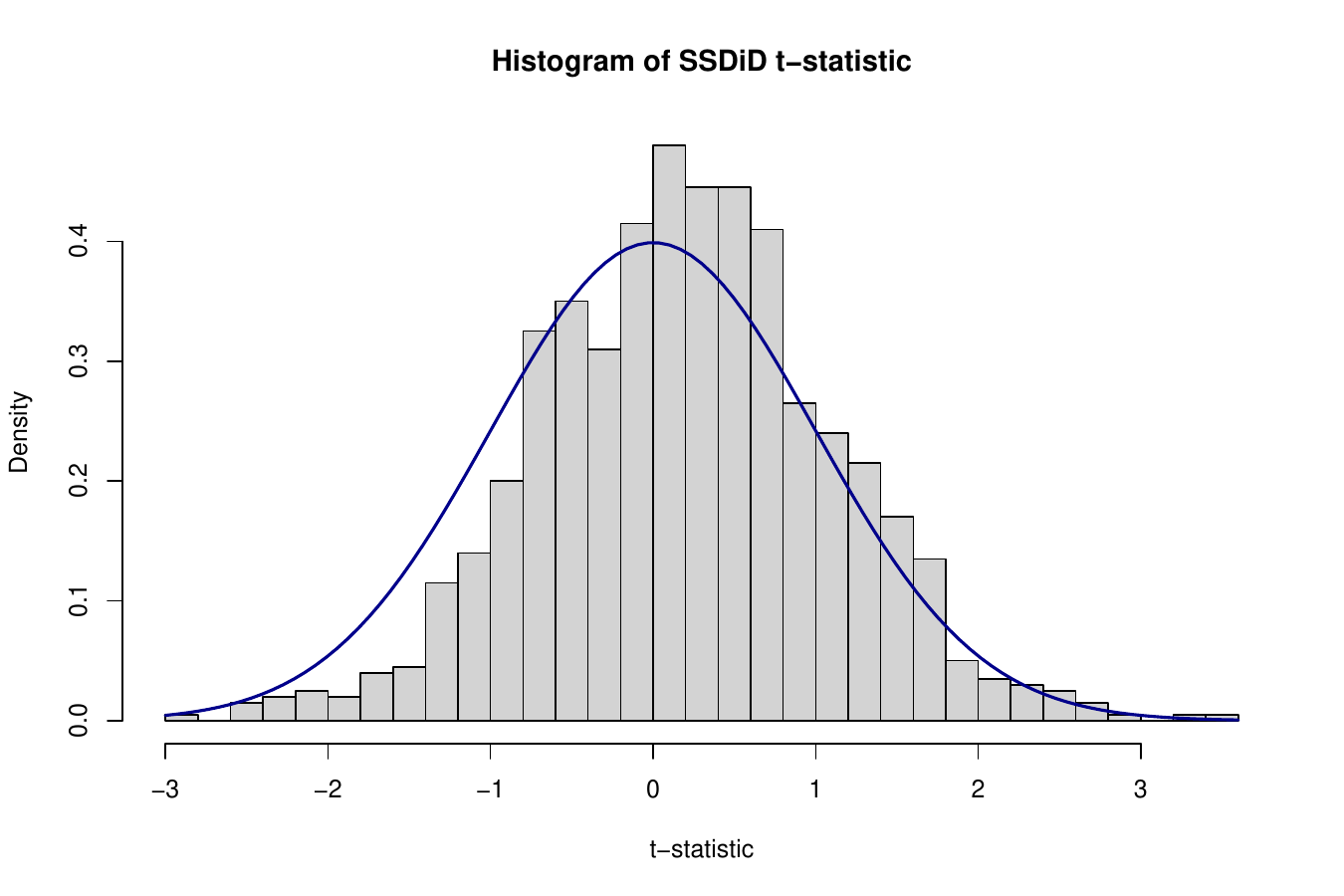}
        \par\vspace{0.5em}
        \small (d) Sequential SDiD, Fifth-Period Effect ($k=4$)
    \end{minipage}

    \vspace{1em}

    \parbox{\textwidth}{\footnotesize \textit{Notes:} The histograms show the distribution of t-statistics over 1000 simulations. Panels (a) and (c) show results for the standard DiD estimator; Panels (b) and (d) show results for the Sequential SDiD estimator. The top row (a--b) corresponds to the contemporaneous effect ($k=0$), and the bottom row (c--d) corresponds to the fifth-period effect ($k=4$). The solid blue line is the standard normal density.}
\end{figure}

\begin{table}[t!]
    \centering
    \caption{Simulation Results: RMSE and Coverage}
    \label{tab:rmse_coverage}
    
    \vspace{0.5em} 

    \textit{Panel A: Root Mean Squared Error (RMSE)}
    \par\vspace{0.25em} 
    \begin{tabular}{lrrrrrrrrr}
        \hline
        Lags & 0 & 1 & 2 & 3 & 4 & 5 & 6 & 7 & 8 \\
        \hline
        SSDiD RMSE & 0.03 & 0.04 & 0.04 & 0.04 & 0.04 & 0.05 & 0.05 & 0.05 & 0.05 \\
        DiD RMSE & 0.05 & 0.05 & 0.05 & 0.05 & 0.06 & 0.06 & 0.06 & 0.07 & 0.07 \\
        \hline
    \end{tabular}

    \vspace{1.5em} 

    \textit{Panel B: Coverage of 95\% Confidence Intervals}
    \par\vspace{0.25em}
    \begin{tabular}{lrrrrrrrrr}
        \hline
        Lags & 0 & 1 & 2 & 3 & 4 & 5 & 6 & 7 & 8 \\
        \hline
        SSDiD & 0.97 & 0.96 & 0.97 & 0.98 & 0.96 & 0.96 & 0.96 & 0.96 & 0.96 \\
        DiD & 0.72 & 0.71 & 0.71 & 0.70 & 0.72 & 0.71 & 0.70 & 0.71 & 0.72 \\
        \hline
    \end{tabular}

    \vspace{1em} 

    \parbox{\textwidth}{\footnotesize \textit{Notes:} The table reports results from 1000 simulations. Panel A reports the Root Mean Squared Error (RMSE) for the Sequential SDiD and standard DiD estimators for treatment effects at different lags. Panel B reports the empirical coverage rates of 95\% confidence intervals constructed using bootstrap standard errors.}
\end{table}

The results demonstrate that Sequential SDiD provides valid inference while the standard DiD estimator fails. Figure~\ref{fig:state_t_stat} shows that the $t$-statistics for the standard DiD estimator are biased for both contemporaneous ($k=0$) and lagged ($k=4$) effects, whereas the $t$-statistics for our estimator remain correctly centered at zero. Table~\ref{tab:rmse_coverage} quantifies this finding: Sequential SDiD has a consistently lower Root Mean Squared Error (RMSE) and its 95\% confidence intervals achieve the nominal coverage rate. In contrast, the coverage for the standard DiD estimator falls to around 70\% due to its bias, rendering it unreliable for inference in this setting.

\begin{remark}[Bootstrap and Aggregation in Simulations]\label{rem:sim_details}
We note two technical details regarding this simulation. First, while our formal theory for the bootstrap is derived for individual-level data where the number of units is large, these results demonstrate its strong empirical performance when applied to aggregated units. A formal proof in this context would require a different asymptotic regime where the number of aggregate units is also large. Additionally, when applying our algorithm to data pre-aggregated by geographic units like states, we make a minor adjustment: for states sharing an adoption time, counterfactuals are imputed in parallel for each state before proceeding to the next horizon, preventing them from being used as controls for one another.
\end{remark}

\section{Conclusion}\label{sec:conc}

We propose a new method, Sequential Synthetic Difference-in-Differences (Sequential SDiD), for estimating treatment effects in event studies with a staggered rollout. Our estimator applies the principles of the original SDiD estimator sequentially to aggregated data, using an iterative imputation procedure where estimates for early cohorts inform those for later ones. We establish the estimator's theoretical properties by proving its asymptotic equivalence to an oracle OLS estimator. This result is significant as it delivers, to our knowledge, the first formal efficiency guarantees for a synthetic control-type method. An empirical application and data-based simulations demonstrate that our estimator is competitive with traditional DiD when its assumptions hold and provides robust inference when they fail.

\newpage
\clearpage
\bibliography{myrefs}{}
\bibliographystyle{chicago}

\newpage
\appendix

\section{Proofs}

\subsection{OLS}
\paragraph{Proof of Proposition \ref{pr:OLS_comp}:}
\begin{proof}
We split the proof into three steps. First, we explain why the OLS estimators are uniquely defined. Second, we show the result for the first `diagonal'', i.e., derive the representation for $\hat \tau^{OLS}_{a,0}$. Finally, we prove the induction step to extend this argument to arbitrary $\hat \tau_{a,k}^{OLS}$.
\begin{enumerate}
    \item The argument for existence of $\hat \tau_{a,k}^{OLS}$ follows from the fact that Assumption \ref{as:rank_cond} guarantees that the solution to all balancing problems in Algorithm \ref{alg:ols} exist. As a result, we know that it is possible to construct an unbiased estimator. Since the OLS estimator optimizes over the set of all unbiased estimators, it follows that the solution to the OLS problem also exists, and has to be unique.
    \item     Fix $a$ and consider $\hat \tau_{a,0}^{OLS}$, which has the representation:
    \begin{equation*}
        \hat \tau_{a,a}^{or} = \sum_{j,l}\hat \omega^{OLS}_{j,l} Y_{j,l},
    \end{equation*}
    where by the standard OLS arguments the weights $\hat \omega^{OLS}(a,a)$ solve the following optimization problem:
    \begin{equation}\label{eq:cons_ols}
        \begin{aligned}
           \hat \omega^{OLS} = \argmin_{\omega} \sum_{j,l}\frac{\omega^2_{j,l}}{\pi_j}\\
            \text{subject to:} \sum_{j}\omega_{j,l} = 0, \sum_{l}\omega_{j,l} = 0\\
            \sum_{j}\omega_{j,l}\theta_j = 0, \sum_{l}\omega_{j,l}\psi_l = 0\\
            \omega_{a,a} = 1\\
            \omega_{j,l} = 0 \text{ for $\infty >j\ge l$, $(j,l) \ne (a, a)$}
        \end{aligned}
    \end{equation}
Define weights $\tilde \omega_{j,l} =(\{j = a\} - \{j \ne a\}\tilde \omega_{j}^{(a)}(\{l = a\} - \{l \ne a\} \tilde\lambda_l^{(a,0)}$ and observe that they satisfy the constraints in \eqref{eq:cons_ols}. For $j <l, (j,l) \ne (a,a)$ and $j = \infty$ the first order conditions for problem \eqref{eq:cons_ols} have the following form:
\begin{equation*}
    \frac{\hat\omega_{j,l}}{\pi_j} = \mu_{0,j} + \gamma_{0,l} + (\mu_{1,l})^\top \theta_j +(\gamma_{1,j})^\top \psi_l  
\end{equation*}
By taking the first-order conditions for the balancing problems in Algorithm \ref{alg:ols}, we have:
\begin{equation*}
\begin{aligned}
    \frac{\tilde \omega_j^{(a)}}{\pi_j} = \mu_{0} + (\mu_{1})^\top \theta_j\\
     \tilde \lambda_l^{(a,0)} = \gamma_{0} + (\gamma_{1})^\top \psi_l\\
\end{aligned}
\end{equation*}
Consider $(j,l)$ such that $j \ne a$ and $l \ne a$. Then we have that 
\begin{align*}
    \frac{\tilde \omega_{j,l}}{\pi_j} = \frac{\tilde\omega_j^{(a)}\lambda_l^{(a,0)}}{\pi_j} = \mu_0\gamma_0 + (\mu_0 \gamma_1)^\top\psi_l + (\gamma_0 \mu_1)^\top \theta_j + \theta_j^{\top} (\mu_1 \gamma_1^\top)\psi_l.
\end{align*}
Next consider $j = a$ but $l \ne a$; we then have:
\begin{equation*}
    \frac{\tilde \omega_{j,l}}{\pi_j} = \frac{- \gamma_0 - \gamma_1^\top \psi_l}{\pi_j},
\end{equation*}
and similarly for $j \ne a$ and $l = a$:
\begin{equation*}
     \frac{\tilde \omega_{j,l}}{\pi_j} = - \mu_0 - \mu_1^\top \theta_j,
\end{equation*}
It then follows:
\begin{equation*}
    \frac{\tilde \omega_{j,l}}{\pi_j} = \tilde \mu_{0,j} + \tilde \gamma_{0,j} + \tilde \mu_{1,j}^\top \theta_j + \tilde \gamma_{1,j}^\top \psi_l,
\end{equation*}
where
\begin{align*}
    \tilde\mu_{0,j} = -\frac{\gamma_0}{\pi_a}\{j =a\} + (\mu_0\gamma_0 +(\gamma_0\mu_1)^\top \theta_j) \{j>a\}\\
    \tilde\gamma_{0,l} = -\mu_0\{l =a\} + ((\mu_0\gamma_1)^\top \psi_l) \{l<a\}\\
     \tilde\gamma_{1,j} = (-\frac{\gamma_1}{\pi_a})\{j=a\} + (\gamma_1\mu^\top)\theta_j \{j >a\}\\
     \tilde\mu_{1,l} =(-\mu_1) \{l = a\}
\end{align*}
It follows that $\tilde \omega_{j,l}$ satisfies the first-order conditions, and since the optimization problem is strictly convex, it implies that the weights $\tilde \omega_{j,l}$ are optimal. 
\item Next, suppose we have shown the result for a given value of $k_0$, and want to extend it to a $k_0+1$. To this end, observe that the OLS estimator for $\hat \tau_{a,k}^{OLS}$ is equivalent to the following optimization problem:
\begin{align*}
  \{\hat\alpha^{OLS}_a,\hat\beta^{OLS}_t, \hat\phi^{OLS}_t, \hat\nu^{OLS}_a,\hat\tau^{OLS}_{a,k}\}_{a,t,k, k > k_0} \in  \\
  \argmin_{\{\alpha_a,\beta_t, \phi_t, \nu_a,\tau_{a,k}\}_{a,t,k}}\sum_{a,t}\pi_a\Bigg(Y_{a,t}  - \{a \le t, t-a \le k_0\}\hat\tau^{OLS}_{a,t-a} - \\
  - \alpha_a - \beta_t - \theta_a^\top \phi_t - \nu_a\psi_t - \{a \le t, t-a > k_0\}\tau_{a,t-a}\Bigg)^2
\end{align*}
This problem has the same structure as the one previous one as soon as we appropriately redefine the outcomes. As a result, we can repeat the same argument as before to prove the induction step.
\end{enumerate}
\end{proof}
\subsection{Abstract quadratic balancing problems}

\subsubsection{Connection I}
Consider two optimization problems:
\begin{align*}
    x^{\star} = \argmin_{x \in L} \| A x-  b\|_2^2 + \eta^2 \|x\|_2^2,\\
    \hat x =  \argmin_{x \in L} \| \hat A x-  b\|_2^2 + \eta^2 \|x\|_2^2
\end{align*}
where $L$ is a convex set. Optimality conditions for the first problem guarantee:
\begin{equation*}
    (A\delta)^\top (Ax^{\star} - b) + 2\eta^2\delta^\top x^{\star} \ge 0
\end{equation*}
for any $\delta \in L - x^{\star}$.

Let $E := \hat A - A$ and $\hat \delta := \hat x - x^{\star}$, then we have from the optimality of $\hat x$:
\begin{align*}
   0 \ge \| \hat A\hat x - b\|_2^2 - \| \hat A x^{\star} - b\|_2^2 + \eta^2(\|\hat x\|_2^2 - \|x^{\star}\|_2^2)= \\
   \| \hat A\hat \delta \|_2^2 +2 (\hat A\hat \delta)^\top (\hat A x^{\star} -b) + \eta^2\left( \|\hat \delta\|_2^2 +2\hat \delta^\top x^{\star}\right) \ge\\
   \|A \hat \delta\|_2^2 + \|E \hat \delta\|_2^2 + 2(E\hat \delta)^\top A\hat \delta + 2(A \hat \delta)Ex^{\star} + 2(E \hat \delta)(Ax^{\star} - b) + 2 (E\hat \delta)^\top Ex^{\star} + \eta^2 \|\hat \delta\|_2^2 \ge \\
\|A \hat \delta\|_2^2-2 \|E\|_{op}\|\hat \delta\|_2 \|A\hat \delta\|_2 +\eta^2 \|\hat \delta\|_2^2  - 2\|A \hat \delta\|\|E\|_{op}\|x^{\star}\|\\
\left(\|E \hat \delta\|_2   - \left(\|Ax^{\star} - b\|_2+\|E\|_{op}\|x^{\star}\|_2\right)\right)^2  -  \left(\|Ax^{\star} - b\|_2+\|E\|_{op}\|x^{\star}\|_2\right)^2.
\end{align*}
We also have the following inequality:
\begin{align*}
    \frac{1}{4} \|A \hat \delta\|_2^2 - 2 \|E\|_{op}\|\hat \delta\|_2 + 4\|E\|_{op}^2 \|\hat \delta\|_2^2 = \left(\frac{1}{2} \|A \hat \delta\|_2 - 2 \|E\|_{op} \|\hat \delta\|_2\right)^2 \ge 0 \Rightarrow\\
    \|A \hat \delta\|_2^2-2 \|E\|_{op}\|\hat \delta\|_2 \|A\hat \delta\|_2 +\eta^2 \|\hat \delta\|_2^2 - 2\|A \hat \delta\|\|E\|_{op}\|x^{\star}\| \ge \\
    \frac{3}{4}  \|A \hat \delta\|_2^2 - 2\|A \hat \delta\|\|E\|_{op}\|x^{\star}\|_2 + (\eta^2  - 4 \|E\|_{op}^2) \|\hat \delta\|_2^2 = \\
     \frac{3}{4}\left( \|A \hat \delta\|_2 - \frac{4}{3}\|E\|_{op}\|x^{\star}\|_2\right)^2 - \frac{4}{3}\|E\|^2_{op}\|x^{\star}\|_2^2 + (\eta^2  - 4 \|E\|_{op}^2) \|\hat \delta\|_2^2.
\end{align*}
Combining all these pieces together, we get:
\begin{align*}
    \frac{3}{4}\left( \|A \hat \delta\|_2 - \frac{4}{3}\|E\|_{op}\|x^{\star}\|_2\right)^2 + (\eta^2  - 4 \|E\|_{op}^2) \|\hat \delta\|_2^2 +\\
    \left(\|E \hat \delta\|_2   - \left(\|Ax^{\star} - b\|_2+\|E\|_{op}\|x^{\star}\|_2\right)\right)^2 \le     \left(\|Ax^{\star} - b\|_2+\|E\|_{op}\|x^{\star}\|_2\right)^2 + \frac{4}{3}\|E\|^2_{op}\|x^{\star}\|_2^2.
\end{align*}
As a result, on the event $\eta^2  - 8 \|E\|_{op}^2> 0$ we get the following implication:
\begin{align*}
    \|A \hat \delta\|_2 \lesssim \|E\|_{op}\|x^{\star}\|_2 + \|Ax^{\star} - b\|_2,\\
    \|E \hat \delta\|_2 \lesssim \|E\|_{op}\|x^{\star}\|_2 + \|Ax^{\star} - b\|_2,\\
    \|\hat \delta\|_2 \lesssim\frac{\|E\|_{op}\|x^{\star}\|_2 + \|Ax^{\star} - b\|_2}{\eta}
\end{align*}

\subsubsection{Connection II}

Consider two optimization problems:
\begin{align*}
    x^{or} = \argmin_{x \in L} \| A x-  b^{or}\|_2^2 + \eta^2 \|x\|_2^2,\\
   x^{\star} =  \argmin_{x \in L} \|  A x-  b\|_2^2 + \eta^2 \|x\|_2^2
\end{align*}
where $L$ is a convex set. Optimality conditions for the first problem guarantee:
\begin{equation*}
    (A\delta)^\top (Ax^{or} - b^{or}) + 2\eta^2\delta^\top x^{or} \ge 0
\end{equation*}
for any $\delta \in L - x^{or}$. Denote $\epsilon := b -b^{or}$ and $\delta^{\star} := x^{\star} - x^{or}$. Using optimality for the second problem, and the optimality conditions for the second problem we get:
\begin{align*}
    0 \ge  \|  A x^{\star}-  b\|_2^2 -\|  A x^{or}-  b\|_2^2 + \eta^2 (\|x^{\star}\|_2^2 -\|x^{or}\|_2^2) = \\
    \|A\delta^{\star}\|^2_2 + 2(A\delta^{\star})^\top(Ax^{or} - b) + \eta^2 (\|\delta^{\star}\|_2^2 + 2(\delta^{\star})^\top x^{or}) \ge \\
     \|A\delta^{\star}\|^2_2 - 2(A\delta^{\star})^\top\epsilon + \eta^2 \|\delta^{\star}\|_2^2 \ge \\
     \|A\delta^{\star}\|^2_2 - 2\|A\delta^{\star}\|_2\|\epsilon\|_2 + \eta^2 \|\delta^{\star}\|_2^2
\end{align*}
It then follows:
\begin{equation*}
     \|A\delta^{\star}\|_2 \lesssim \|\epsilon\|_2, \quad \|\delta^{\star}\|_2 \lesssim \frac{ \|\epsilon\|_2}{ \eta}
\end{equation*}
Combining the results of the two problems together, we can conclude that on the event $\eta^2  - 8 \|E\|_{op}^2> 0$, we get the following bounds
\begin{align*}
    \|A (\hat x - x^{or})\|_2 \lesssim  \|E\|_{op}\left(\|x^{or}\|_2 +\frac{ \|\epsilon\|_2}{ \eta}\right) + \|Ax^{or} - b^{or}\|_2 +  \|\epsilon\|_2\\
    \| \hat x - x^{or}\|_2 \lesssim \frac{\|E\|_{op}\left(\|x^{or}\|_2 +\frac{ \|\epsilon\|_2}{ \eta}\right) + \|Ax^{or} - b^{or}\|_2  + \|\epsilon\|_2}{\eta}
\end{align*}

\subsubsection{Connection III}
Consider a vector $y$ that satisfies:
\begin{equation*}
    Ay = b^{or}, \quad c^\top y = 1.
\end{equation*}
We use $c^\top y = 1$ in place of $L$ in the previous discussion and write
\begin{equation*}
    x^{or} := \arg\min_{c^\top x = 1}\Bigl\{\| A x - b^{or}\|_2^2 + \eta^2 \| x\|_2^2\Bigr\}.
\end{equation*}
Using strong duality, we get the following equivalence:
\begin{multline*}
    \min_{c^\top x = 1}\Bigl\{\| A x - b^{or}\|_2^2 + \eta^2 \| x\|_2^2\Bigr\} = \min_{x,t}\max_{\|\beta\|_2\le 1, \beta_0, \lambda\ge 0}\Bigl\{ \lambda( \beta^\top(Ax - b^{or}) - t) + t^2 + \eta^2 \|x\|_2^2 +\\
    \beta_0(c^\top x-1)\Bigr\} = \max_{\beta, \beta_0}\min_{x}\Bigl\{ \beta^\top(Ax - b^{or})  -\frac{\|\beta\|_2^2}{4} + \eta^2 \|x\|_2^2 + \beta_0(c^\top x-1)\Bigr\}= \\
    - \eta^2 \min_{\beta, \beta_0} \Bigl\{\left\|A^\top \beta + \beta_0c\right\|_2^2+\eta^2\|\beta\|_2^2 - 2\left(A^\top \beta + \beta_0c\right)^\top y\Bigr\} = \\
     - \eta^2 \min_{\beta, \beta_0} \Bigl\{\left\|y - A^\top \beta - \beta_0c\right\|_2^2+\eta^2\|\beta\|_2^2  - \|y\|_2^2\Bigr\},
\end{multline*}
where the solution to the primal problem satisfies
\begin{equation*}
\begin{aligned}
        (\beta^{\star},\beta_0^{\star}) &= \arg\min_{\beta, \beta_0} \Bigl\{\left\|y - A^\top \beta - \beta_0c\right\|_2^2+\eta^2\|\beta\|_2^2\Bigr\},\\
        x^{or} &=A^\top \beta^{\star}+ \beta_0c^{\star}.
\end{aligned}
\end{equation*}
Using the first-order conditions for the dual problem, we get:
\begin{equation*}
    b^{or} - Ax^{or} = A (y - x^{or}) = A(y - A^\top \beta^{\star} - \beta_0^{\star}c) = \eta^2 \beta^{\star} \Rightarrow \|b^{or} - Ax^{or}\|_2 = \eta^2 \|\beta^{\star}\|_2
\end{equation*}
This implies that the imbalance $\|b - Ax^{or}\|_2$ is of the order of $ \eta^2 $ as long as $\|\beta^{\star}\|_2$ is bounded. Observe that this bound is much better than the trivial bound
\begin{equation*}
    \| A x^{or} - b^{or}\|_2^2 \le \eta^2 \|y\|_2^2,
\end{equation*}
which we get directly from the optimality of $x^{or}$ and properties of $y$.

Observe that the result above holds for any $y$ that satisfies the equations. To bound $\beta^{\star}$ we consider a least-norm solution:
\begin{equation*}
 \begin{aligned}
     &y^{or} := \argmin_{y} \|y\|_2^2\\
     \text{subject to: } &Ay = b^{or}, \quad c^\top y = 1.
 \end{aligned}
\end{equation*}
Solution to this problem has the form $y^{or} = A^\top \beta^{or} + c \beta^{or}_0$, and by construction $\| \beta^{or}\|_2 \ge \|\beta^{\star}\|_2$. As a result, we can bound the imbalance
\begin{equation*}
    \|b^{or} - A x^{or}\|_2 \le \eta^2 \| \beta^{or}\|_2.
\end{equation*}
We also have $\|x^{or}\|_2 \le \| y^{or}\|_2$. We also use the following bound
\begin{align*}
    \|y^{or}\|_2 = \| A^\top \beta^{or} + c \beta^{or}_0\|_2 \ge \|P_{c^\bot} A^\top \beta^{or}\|_2 \ge \| P_{c^\bot}A\|_{\min} \|\beta^{or}\|_2 \Rightarrow\\
    \|\beta^{or}\|_2\le \frac{\|y^{or}\|_2}{\|P_{c^\bot} A\|_{\min}},
\end{align*}
where $P_{c^\bot}$ is the orthogonal projector on the complement of $c$ and $\| P_{c^\bot}A\|_{\min}$ is the smallest singular value of $\|P_{c^\bot}A\|_{\min}$. The last result relies on the fact that $\beta^{or}$ has no component in the kernel of the matrix $A^{\top}$. This is without loss of generality because if such components exist, then we can drop it and redefine $\beta^{or}$ without changing the results. The argument also relies on $\|P_{c^\bot} A\|_{\min}$ being non-zero, but this is also without loss of generality, because if it is zero, then we can set $\beta^{or}$ to zero as well.

We also have:
\begin{equation*}
    \| A\delta^{\star}\|_2 \ge \|A\|_{\min}\|\delta^{\star}\|_2 \Rightarrow \|\delta^{\star}\|_2 \le \frac{\eta^2}{\|A\|_{\min}\|P_{c^\bot} A\|_{\min}} \le \frac{\eta^2}{\|P_{c^\bot} A\|^2_{\min}},
\end{equation*}
where we used the fact that $\delta^{\star}$ belongs to the image of $A^\top$.

Combining these results with our previous discussion, we can conclude that on the event $\eta^2  - 8 \|E\|_{op}^2> 0$, we get the following bounds:

\begin{align*}
    \|A (\hat x - y^{or})\|_2 \lesssim  \|E\|_{op}\left(\|y^{or}\|_2 +\frac{ \|\epsilon\|_2}{ \eta}\right) +\frac{\eta^2 \|y^{or}\|_2}{\|P_{c^\bot} A\|_{\min}}  +  \|\epsilon\|_2\\
    \| \hat x - x^{or}\|_2 \lesssim \frac{\|E\|_{op}\left(\|y^{or}\|_2 +\frac{ \|\epsilon\|_2}{ \eta}\right) + \frac{\eta^2 \|y^{or}\|_2}{\|P_{c^\bot} A\|_{\min}}  + \|\epsilon\|_2}{\eta} + \frac{\eta^2\|y^{or}\|_2}{\|P_{c^\bot} A\|^2_{\min}}
\end{align*}

\subsection{Sequential SDiD}
\paragraph{Proof of Theorem \ref{th:as_connection}:}
The argument relies on applying the results established in the previous section. First, we define an error matrix $E$ such that.
\begin{equation*}
    E_{j,l} = \epsilon_{j,l}
\end{equation*}
We also define $E^{(a,k)}$ -- the top-left corner of matrix $E$ that corresponds to periods $t < a+k$ and adoption times $j > a$. We also define $E^{(a,k)}_{a.} = (\varepsilon_{a,1},\dots, \varepsilon_{a,a+k-1})$ and $E^{(a,k)}_{.k} = (\varepsilon_{a+1,a+k},\dots, \varepsilon_{\infty,a+k})^\top$. Assumption \ref{as:var_behavior} guarantees that $\|E^{(a,k)}\|_{op}  = O_p\left(\frac{1}{\sqrt{n}}\right)$ via Markov inequality for all $(a,k)$ and the same hols for $E^{(a,k)}_{.k}$ and $E^{(a,k)}_{a.}$.

The proof is based on the induction argument. We start by analyzing the difference
\begin{equation*}
    \hat \tau_{a,a}^{SSDID}-  \hat \tau_{a,a}^{OLS},
\end{equation*}
establish the rate for this difference, assume that the same rate holds for $\hat \tau_{a,a+k}^{SSDID}-  \hat \tau_{a,a+k}^{OLS}$ and finally prove the induction step by showing that it implies the same rate for $\hat \tau_{a,a+k+1}^{SSDID}-  \hat \tau_{a,a+k+1}^{OLS}$. 

\begin{enumerate}
    \item   We start with the expansion:
    \begin{align*}
            \hat\tau^{SSDID}_{a,a} - \hat\tau^{OLS}_{a,a} =  \left(Y_{a,a} - \left(Y^{(a,a)}_{a.}\right)\hat \lambda^{(a,0)}\right)  - 
            \left((\hat \omega^{(a,a)})^\top Y^{(a,a)}_{.0}  - (\hat \omega^{(a,a)})^\top Y^{(a,a)}\hat \lambda^{(a,0)}\right)\\
            -\left(Y_{a,a} - \left(Y^{(a,a)}_{a.}\right)\tilde \lambda^{(a,0)}\right)  +
            \left((\tilde \omega^{(a)})^\top Y^{(a,a)}_{.0}  - (\tilde \omega^{(a)})^\top Y^{(a,a)}\tilde \lambda^{(a,0)}\right)
            \\
            = (\hat \delta^{(a,a)}_{\omega})^{\top} Y^{(a,a)}\hat \delta^{(a,a)}_{\lambda} + 
            (\tilde \omega^{(a)})^\top Y^{(a,a)}\hat \delta^{(a,a)}_{\lambda} +(\hat \delta^{(a,a)}_{\omega})^{\top} Y^{(a,a)}\tilde \lambda^{(a,0)} -
            Y_{a.}^{(a,a)}\hat \delta^{(a,a)}_{\lambda}  - (\hat \delta^{(a,a)}_{\omega})^{\top} Y^{(a,a)}_{.0} =
            \\
           =(\hat \delta^{(a,a)}_{\omega})^{\top} L^{(a,a)}\hat \delta^{(a,a)}_{\lambda} + 
            (\tilde \omega^{(a)})^\top L^{(a,a)}\hat \delta^{(a,a)}_{\lambda} +(\hat \delta^{(a,a)}_{\omega})^{\top} L^{(a,a)}\tilde \lambda^{(a,0)} -
            L^{(a,a)}_{a.}\hat \delta^{(a,a)}_{\lambda}  - (\hat \delta^{(a,a)}_{\omega})^{\top} L_{.0}+ \\
            (\hat \delta^{(a,a)}_{\omega})^{\top} E^{(a,a)}\hat \delta^{(a,a)}_{\lambda} + 
            (\tilde \omega^{(a)})^\top E^{(a,a)}\hat \delta^{(a,a)}_{\lambda} +(\hat \delta^{(a,a)}_{\omega})^{\top} E^{(a,a)}\tilde \lambda^{(a,0)} -
            E^{(a,a)}_{a.}\hat \delta^{(a,a)}_{\lambda}  - (\hat \delta^{(a,a)}_{\omega})^{\top} E^{(a,a)}_{.0}\le\\
             (\hat \delta^{(a,a)}_{\omega})^{\top} L^{(a,a)}\hat \delta^{(a,a)}_{\lambda} +\\
             \| E^{(a,a)}\|_{op}\|\|\hat\delta^{(a,a)}_{\omega})\|_2 \|\hat \delta^{(a,a)}_{\lambda}\|_2 +  \| E^{(a,a)}\|_{op}\|\hat\delta^{(a,a)}_{\omega})\|_2\|\tilde \lambda^{(a,0)}\| + \| E^{(a,a)}\|_{op}\|\tilde \omega^{(a)}\|_2 \|\hat \delta^{(a,a)}_{\lambda}\|_2 + \\
              \|\hat \delta^{(a,a)}_{\lambda}\|_2\| E^{(a,a)}_{.0}\|_2 + \|\hat \delta^{(a,a)}_{\lambda}\|_2 \| E^{(a,a)}_{a.}\|_2
        \end{align*}
    Where we used the fact $L^{(a,a)}\tilde \lambda^{(a,0)} = L^{(a,a)}_{a.}$ and similarly $(\tilde \omega^{(a)})^{\top}L^{(a,a)} = L^{(a,a)}_{.0}$ by the definition of the OLS weights. We can also decompose:
    \begin{equation*}
        | (\hat \delta^{(a,a)}_{\omega})^{\top} L^{(a,a)}\hat \delta^{(a,a)}_{\lambda}| \le \frac{\| (\hat \delta^{(a,a)}_{\omega})^{\top} L^{(a,a)}\|_2\|L^{(a,a)}\hat \delta^{(a,a)}_{\lambda}\|_2}{\tilde \sigma_{a,a}}.
    \end{equation*}
    We can now apply the abstract balancing bounds established in the previous section to conclude:
\begin{align*}
    \|L^{(a,a)} \hat \delta^{(a,a)}_{\lambda}\|_2 &\lesssim_p  \frac{1}{\sqrt{n}}\left(\|\tilde \lambda^{(a,0)}\|_2 +\frac{1}{ \sqrt{n}\eta}\right) +\frac{\eta^2\|\tilde \lambda^{(a,0)}\|_2}{\tilde \sigma_{a,a}}  + \frac{1}{\sqrt{n}},\\
    \| \hat \delta^{(a,a)}_{\lambda}\|_2 &\lesssim_p \frac{\frac{1}{\sqrt{n}}\left(\|\tilde \lambda^{(a,0)}\|_2 +\frac{1}{ \sqrt{n}\eta}\right) +\frac{\eta^2\|\tilde \lambda^{(a,0)}\|_2}{\tilde \sigma_{a,a}}  + \frac{1}{\sqrt{n}}}{\eta} + \frac{\eta^2\|\tilde \lambda^{(a,0)}\|_2}{\tilde \sigma^2_{a,a}},\\
        \|(\hat \delta^{(a,a)}_{\omega})^{\top}L^{(a,a)} \|_2 &\lesssim_p  \frac{1}{\sqrt{n}}\left(\|\tilde \omega^{(a)}\|_2 +\frac{1}{ \sqrt{n}\eta}\right) +\frac{\eta^2\|\tilde \omega^{(a)}\|_2}{\tilde \sigma_{a,a}}  + \frac{1}{\sqrt{n}},\\
    \|\hat \delta^{(a,a)}_{\omega}\|_2 &\lesssim_p \frac{\frac{1}{\sqrt{n}}\left(\|\tilde \omega^{(a)}\|_2 +\frac{1}{ \sqrt{n}\eta}\right) +\frac{\eta^2\|\tilde \omega^{(a)}\|_2}{\tilde \sigma_{a,a}}  + \frac{1}{\sqrt{n}}}{\eta} + \frac{\eta^2\|\tilde \omega^{(a)}\|_2}{\tilde \sigma^2_{a,a}}.
\end{align*}
By assumption the variance of the OLS estimator is finite, which together with Assumption \ref{as:var_behavior} guarantees that $\|\tilde \omega^{(a)}\|_2 \sim 1$ and $\|\tilde \lambda^{(a,0)}\|_2 \sim1 $ (the lower bound follows from the fact that they sum up to 1). Putting these results together we can conclude:
\begin{align*}
    |\hat\tau^{SSDID}_{a,a} - \hat\tau^{OLS}_{a,a}| \lesssim_{p} \frac{\left(\frac{1}{\sqrt{n}} + \frac{\eta^2}{\tilde \sigma_{a,a}}\right)^2}{\tilde \sigma_{a,a}} + \frac{1}{\sqrt{n}}\left( \frac{1}{\sqrt{n}\eta} + \frac{\eta}{\tilde \sigma_{a,a}}\right) \lesssim_p \\
    \frac{1}{n\tilde \sigma_{a,a}} + \eta\left(\frac{\eta}{\tilde \sigma_{a,a}}\right)^3 + \frac{1}{\sqrt{n}}\left( \frac{1}{\sqrt{n}\eta} + \frac{\eta}{\tilde \sigma_{a,a}}\right) \ll_{p}n^{-\frac12}.
\end{align*}
We can then repeat this argument for all feasible $a$.
\item We now establish the induction step. We have the following:
\begin{align*}
          \hat\tau^{SSDID}_{a,k} - \hat\tau^{OLS}_{a,k} =  \left(Y_{a,a} - \left(Y^{(a,a)}_{a. } - \hat\tau^{SSDID,(a,k)}_{a.}\right)\hat \lambda^{(a,0)}\right)  - \\
          \Bigg((\hat \omega^{(a,a)})^\top (Y^{(a,a)}_{.0} -\hat\tau^{SSDID,(a,k)}_{.0}) -  (\hat \omega^{(a,a)})^\top (Y^{(a,a)}- \hat\tau^{SSDID,(a,k)})\hat \lambda^{(a,0)}\Bigg) - \\ 
            -\left(Y_{a,a} - \left(Y^{(a,a)}_{a.} - \hat\tau^{OLS,(a,k)}_{a.}\right)\tilde \lambda^{(a,0)}\right)  +\\
            \left((\tilde \omega^{(a)})^\top (Y^{(a,a)}_{.0} -\hat\tau^{OLS,(a,k)}_{.0}) - (\tilde \omega^{(a)})^\top (Y^{(a,a)} - \hat\tau^{OLS,(a,k)})\tilde \lambda^{(a,0)}\right)=\\
            \text{part 1} + \text{part 2} + \text{part 3},
\end{align*}
where
\begin{align*}
    \text{part 1} := &(\hat \delta^{(a,k)}_{\omega})^{\top} (Y^{(a,k)} - \tau^{(a,k)})\hat \delta^{(a,k)}_{\lambda} + 
            (\tilde \omega^{(a)})^\top (Y^{(a,k)}-\tau^{(a,k)})\hat \delta^{(a,k)}_{\lambda} +\\&(\hat \delta^{(a,k)}_{\omega})^{\top} (Y^{(a,k)} - \tau^{(a,k)})\tilde \lambda^{(a,k)} -
            (Y^{(a,k)}_{a.}-\tau^{(a,k)}_{a.})\hat \delta^{(a,k)}_{\lambda}  - (\hat \delta^{(a,k)}_{\omega})^{\top} (Y^{(a,k)}_{.0} - \tau^{(a,k)}_{.0});\\
            \text{part 2} := &(\hat \delta^{(a,k)}_{\omega})^{\top} (\tau^{OLS,(a,k)} - \tau^{(a,k)})\hat \delta^{(a,k)}_{\lambda} + 
            (\tilde \omega^{(a)})^\top (\tau^{OLS,(a,k)}-\tau^{(a,k)})\hat \delta^{(a,k)}_{\lambda} +\\
            &(\hat \delta^{(a,k)}_{\omega})^{\top} (\tau^{OLS,(a,k)} - \tau^{(a,k)})\tilde \lambda^{(a,k)} -
            (\tau^{OLS,(a,k)}_{a.}-\tau^{(a,k)}_{a.})\hat \delta^{(a,k)}_{\lambda}  - (\hat \delta^{(a,k)}_{\omega})^{\top} (\tau^{OLS,(a,k)}_{.0} - \tau^{(a,k)}_{.0});\\
                \text{part 3} :=&\left(\hat\tau^{SSDID,(a,k)}_{a.} - \hat\tau^{OLS,(a,k)}_{a.}\right)\hat \lambda^{(a,0)}  + \\
          &\Bigg((\hat \omega^{(a,a)})^\top (\hat\tau^{SSDID,(a,k)}_{.0} - \hat\tau^{OLS,(a,k)}_{.0}) -  (\hat \omega^{(a,a)})^\top (\hat\tau^{SSDID,(a,k)} - \hat\tau^{OLS,(a,k)})\hat \lambda^{(a,0)}\Bigg).
\end{align*}
The induction assumption and the fact that the weights are bounded guarantees that the last terms is $o_p\left(n^{-\frac12}\right)$. To establish the bounds for the first two parts we need to guarantee that we have the same guarantees for the weights error as before. It is easy to see that this is the case, though, because the OLS etimator has  errors of the order $\frac{1}{\sqrt{n}}$, and the deviations of the Sequential SDiD estimator from the OLS estimator are of the smaller order by induction assumption. As a result, we get the same bounds as before, and can guarantee that both parts are of the order  $\frac{1}{\sqrt{n}}$, thus concluding the proof. 
\end{enumerate}

\paragraph{Inference}
The asymptotic equivalence established in Theorem \ref{th:as_connection} can be shown to hold for the corresponding bootstrap estimators. When the Bayesian bootstrap is applied, the aggregate data points become weighted averages, with weights $\{\xi_i\}_{i=1}^n$ drawn from an i.i.d. Exponential distribution. Consequently, the error matrix $E$ in the proof is replaced by its bootstrap analog, $E(\xi)$. Because the bootstrap weights have finite moments, the operator norm of $E(\xi)$ can be shown to converge at the same rate as that of $E$, conditional on the data.

Therefore, the arguments in the proof of Theorem \ref{th:as_connection} apply to the bootstrapped estimators, which implies that
$$
\hat{\tau}_{a,k}^{SSDiD}(\xi) = \hat{\tau}_{a,k}^{OLS}(\xi) + o_p(n^{-1/2})
$$
conditional on the data. This equivalence allows us to use the bootstrap distribution of our estimator for inference, as its validity is inherited from the known validity of the Bayesian bootstrap for OLS estimators (e.g., \citealp{chamberlain2003nonparametric}).

\end{document}